\documentclass[11pt,a4paper]{article}
\pdfoutput=1
\usepackage{jheppub}
\usepackage{amsmath,amsfonts}
\usepackage{multicol}
\usepackage{graphicx}
\usepackage{float}
\usepackage{hyperref}
\usepackage[plain]{fancyref}
\usepackage{array}
\usepackage{caption}
\usepackage{subcaption}
\usepackage{xspace}
\def\beq{\begin{equation}}
\def\eeq{\end{equation}}

\let\ep=\epsilon

\let\ka=\kappa

\let\si=\sigma

\newcommand{\erf}{\text{erf}}
\newcommand{\kev}{\text{keV}}
\newcommand{\kdn}{$k$-DN\xspace}
\newcommand{\dm}{dark matter\xspace}
\newcommand{\nreq}{$N_\text{req}$\xspace}
\newcommand{\nmax}{$N_\text{max}$\xspace}

\newcommand{\ie}{{\it i.e }}

\newcolumntype{M}[1]{>{\centering\arraybackslash}m{#1}}
\newcolumntype{L}[1]{>{\arraybackslash}m{#1}}
\newcolumntype{N}{@{}m{0pt}@{}}

\newcommand{\deap}{DEAP-3600\xspace}
\newcommand{\xenon}{XENON1T\xspace}

\title{Can Tonne-Scale Direct Detection Experiments Discover Nuclear Dark Matter?} 

\author{Alistair~Butcher,}
\author{Russell~Kirk,}
\author{Jocelyn~Monroe}
\author{and Stephen~M.~West}

\affiliation{Dept.\ of Physics,
Royal Holloway University of London, Egham, Surrey, TW20 0EX, U.K.}

\emailAdd{Alistair.Butcher.2010@live.rhul.ac.uk}
\emailAdd{Russell.Kirk.2008@live.rhul.ac.uk}
\emailAdd{Jocelyn.Monroe@rhul.ac.uk}
\emailAdd{Stephen.West@rhul.ac.uk}

\abstract{ Models of nuclear dark matter propose that the dark sector contains large composite states consisting of dark nucleons in analogy to Standard Model nuclei. We examine the direct detection phenomenology of a particular class of nuclear dark matter model at the current generation of tonne-scale liquid noble experiments, in particular \deap and \xenon. In our chosen nuclear dark matter scenario distinctive features arise in the recoil energy spectra due to the non-point-like nature of the composite dark matter state. We calculate the number of events required to distinguish these spectra from those of a standard point-like WIMP state with a decaying exponential recoil spectrum. In the most favourable regions of nuclear dark matter parameter space, we find that a few tens of events are needed to distinguish nuclear dark matter from WIMPs at the $3\,\si$ level in a single experiment. Given the total exposure time of \deap and \xenon we find that at best a $2\,\si$ distinction is possible by these experiments individually, while $3\,\si$ sensitivity is reached for a range of parameters by the combination of the two experiments. We show that future upgrades of these experiments have potential to distinguish a large range of nuclear dark matter models from that of a WIMP at greater than $3\,\si$. 
}

\begin{document}
\maketitle
\flushbottom

\section{Introduction}\label{sec:Intro}
A large body of indirect evidence has strongly motivated the existence of \dm; non-baryonic matter constituting approximately a quarter of the Universe's energy density \cite{Ade:2015xua}. In order to fully explain \dm, we are compelled to consider Beyond the Standard Model (BSM) theories as the Standard Model (SM) is incapable of accounting for all of the \dm energy density. Most BSM theories assume that \dm takes the form of (or at least interacts as) point-like particles, but this is not necessarily the case. The SM already provides a clear example of a more complicated structure with normal matter existing in a rich spectrum of composite states. We are therefore motivated to ask whether a similar structure exists in the dark sector and whether this leads to important phenomenological consequences.

The idea that the \dm states in our Universe are not point-like is not new, for example the formation of bound states of two WIMP particles was considered in \cite{Pospelov2008,MarchRussell2008,Shepherd2009}, and a recent analysis has examined the generic direct detection phenomenology of these WIMPonium states~\cite{Laha:2013gva,Laha:2015yoa}. It is also possible to build dark analogues of simple SM atoms\footnote{For related and earlier attempts to explain \dm using a SM analogue see for example \cite{Hodges:1993yb,Goldberg:1986nk,Berezhiani:1995yi,Berezhiani:1995am,Foot:1995pa,Mohapatra:2000qx,Foot:2004pa}.}, for example dark Hydrogen \cite{Kaplan:2009de}, consisting of a bound state formed between a dark proton and a dark electron. This idea has also been extended to the case of dark molecular objects leading to interesting limits on their atomic properties from constraints on their self-scattering cross-sections \cite{Cline:2013pca}.

This paper focuses on the possibility of nuclear dark matter (NDM), where \dm is composed of bound states of strongly-interacting dark nucleon (DN) constituents with short-range interactions \cite{Krnjaic2014,Detmold2014,Wise:2014jva,Wise:2014ola,Hardy:2014mqa,Hardy:2015boa}. In particular, we will consider the class of dark nuclei outlined in \cite{Hardy:2014mqa,Hardy:2015boa} focussing on the potential signals in (and constraints from) direct detection experiments. This class of model assumes that the dark nuclei have an approximately uniform density, with a hard core repulsion between the constituent DNs. This is in contrast to the \dm ``nuggets" described in \cite{Wise:2014jva,Wise:2014ola}, where no hard core repulsion was assumed resulting in a more complicated internal structure. 

In comparison to standard WIMP \dm, the mass of a typical dark nucleus can be much larger. For example, in \cite{Hardy:2014mqa,Hardy:2015boa} dark nuclei with a large number of constituents each with masses in the GeV region were considered leading to states with a total mass well in excess of the unitarity limit of thermal WIMP \dm.

The composite and extended nature of these \dm states will affect the way in which they scatter off SM nuclei in direct detection experiments. If all \dm is in the form of composite states each with $k$ dark nucleons (which we refer to as a \kdn state), and if the momentum transfer in the scattering process is less than the inverse radius of the state then the elastic scattering cross-section will be coherently enhanced by a factor of $k^2$ \cite{Gelmini:2002ez,Krnjaic2014,Hardy:2014mqa,Hardy:2015boa}. However, as we increase $k$ the number density of these \kdn states will decrease, leaving an overall event rate increasing linearly with $k$ \cite{Hardy:2015boa}.

The dependence of the scattering rate on the spatial properties of both the extended \dm states and the SM nuclei is encoded in the product of their respective form factors. These form factors are as usual related to the Fourier transforms of the spatial densities of the particular composite object. If the radii of the \dm states are larger than the SM nuclei, off which they are scattering, then direct detection experiments will start to probe the dark form factor at lower values of the momentum transfer compared to the SM form factor. This can lead to a striking structure in the recoil spectrum \cite{Hardy:2015boa}. For example, if we assume that the composite \dm states possess a uniform internal density, and model this density in terms of a spherical top hat function, the resulting form factor is a spherical Bessel function. The recoil energy spectrum will as a result include characteristic peaks and troughs. Whether these features are distinguishable depends upon the individual experiment's energy resolution and threshold, and is the focus of this work. 

The appearance of these distinctive peaks and troughs in the recoil spectrum also depends on whether all dark matter states have the same number of DNs or whether there is a distribution of nuclei with different $k$ values. The position of the troughs and peaks in the recoil spectrum is dependent on the value of $k$ and it was shown in \cite{Hardy:2015boa} that the compound recoil spectrum from a distribution of $k$ values results in the peaks and troughs being smoothed out leaving a monotonically decreasing spectrum. While more difficult, in a single detector with a sufficiently high number of signal events the recoil spectrum in this case may still be distinguishable from that of a standard WIMP for a given \dm velocity distribution. Due to the uncertainty in the \dm velocity distribution a more effective way to distinguish between NDM and WIMP spectra is to perform a halo-independent analysis using data from several different detectors (see for example \cite{Cherry:2014wia}). In this paper however we wish to study the simpler case of dark nuclei existing predominantly with a fixed number of DNs leaving the halo-independent analysis for future work. 

In \cite{Hardy:2015boa} a simplified direct detection analysis of fixed $k$ dark nuclei was performed using approximations of future detectors. In this paper we build on this work and perform an in-depth analysis of NDM at \deap \cite{Amaudruz:2014nsa} and \xenon \cite{Aprile:2015uzo} using the predicted detectors' energy response functions, efficiencies and thresholds. 

The first detector we consider is the liquid argon detector \deap, which will soon begin its 3 year run and will be roughly an order of magnitude more sensitive to the WIMP scattering cross-section than current searches at a 100 GeV WIMP mass. We have chosen \deap as one of our test cases as it has excellent energy resolution, which will be important if we hope to be able to identify peaks and troughs in the recoil energy spectrum. \deap also has a better sensitivity to high mass dark matter states due to the slow (compared to xenon) fall off in the argon nuclear form factor. 

The second direct detection experiment \xenon is a liquid xenon detector which will also soon begin its physics run, and will be roughly two orders of magnitude more sensitive to the WIMP scattering cross-section than current searches in the mass range of 10-50 GeV. The use of the xenon target allows for a few-keV energy threshold, potentially giving sensitivity to features in the recoil energy spectrum at the lower energies where the scattering rate is largest.

Although our focus is to analyse the potential distinguishability of NDM from a standard WIMP signal, this paper includes an analysis of the current and future constraints on NDM if no signal is observed. We use the limits on WIMP \dm from LUX \cite{Akerib:2015rjg} to calculate the current upper bounds on the DN-SM nucleon scattering cross-section for fixed $k$ dark nuclei and go on to calculate the potential future limits arising from \deap and \xenon. 

Following this introduction, in \Fref{sec:NDM} we outline our chosen model of NDM and review the calculation of the recoil spectrum. In \Fref{sec:Sens} we detail the specifications of \deap, \xenon, and LUX, including the predicted energy resolutions, efficiencies, and energy thresholds used in our analyses. We go on to explore the generic features of the recoil energy spectra for a selection of benchmark NDM scenarios at \deap and \xenon, detailing the effect of the energy resolution and energy threshold on the observability of the features produced by dark form factors. We then present the current and projected (if no signals are observed) limits on NDM from LUX and the two tonne-scale direct detection experiments respectively. In \Fref{sec:Likelihood} we turn to the main component of this work, calculating the number of signal events required in order to distinguish a NDM spectrum from a decaying exponential recoil spectrum produced by a WIMP. In light of this analysis we discuss the potential for the discovery of NDM at \deap and \xenon. Finally we discuss our results and conclude in \Fref{sec:Con}.

\section{Nuclear Dark Matter Recoil Spectrum}\label{sec:NDM}
The scenario we consider here is a variant of the general class of model described in \cite{Hardy:2014mqa}, where dark nuclei generically exist in a wide distribution of sizes. The assumptions of \cite{Hardy:2014mqa} include the existence of a binding energy per DN that saturates (rather than turning over) at large dark nucleon number leading to fusion processes that continue to produce larger and larger dark nuclei with the limit in size being determined by the temperature at which the fusion processes freeze-out. This results in a distribution over different sizes of dark nuclei. In this case, the predicted peaks and troughs in the recoil spectrum expected for a single size of dark nucleus are washed out when the cumulative effect of the wide distribution over $k$ is taken into account \cite{Hardy:2015boa}. 

The modified case where the distribution of dark nuclei sizes are focused around a particular value of $k$ can in principle arise if the binding energy per DN turns over at some nucleon number, in the same way the SM nuclear binding energy per nucleon does at iron. This turnover, for example, could arise in the dark sector in an analogous way to the SM by introducing a dark equivalent of the coulomb force. The direct detection prospects of this scenario is qualitatively different to that of the wide distribution and deserves detailed attention. This is the scenario we consider here, \ie the present \dm abundance is assumed to consist entirely of dark nuclei with same number of DNs. Given the structure in the recoil spectrum for our chosen case, it is likely that this scenario would be the easiest to identify and hence is a sensible case with which to start. 

In this analysis we will consider the underlying interaction between the DNs and SM quarks being due to the exchange of a heavy scalar mediator. The elastic scattering rate of two spatially extended states will then depend on the product of the respective form factors derived from the Fourier transforms of the individual spatial densities of the colliding objects.

The composite \dm states we consider have an approximately uniform density of constituent matter and we model the density as a spherical top hat function, leading to a spherical Bessel function form factor,
\beq\label{eq:DNFormFactor}
F_k(q) = \frac{3 j_1(qR_k)}{qR_k},
\eeq
where $R_k$ is the radius of the dark mater state and $q$ is the momentum transfer in the collision. Under the assumption of a constant density and negligible binding energy, the radius of a \kdn state is approximated by
\beq\label{eq:Rk}
R_k = k^{1/3} R_1,
\eeq
where $R_1$ is the approximate radius of a single DN.

For the SM nuclear form factor, we use the Helm parameterisation \cite{Helm:1956zz}
\beq
F_N(q) = \frac{3 j_1(qR_N)}{qR_N} e^{-q^2 s^2/2},
\eeq
which represents a modification to the spherical top hat density to include a finite-width drop off with skin depth $s = 0.9$\;fm. For the SM nuclei's radius $R_N$, we employ the approximate analytic expression
\beq
R_N^2 = c^2 + \frac{7}{3} \pi^2 a^2 - 5 s^2,
\eeq
with $c \approx 1.23 A^{1/3} - 0.6$ fm (where $A$ is the SM nucleon number), and $a = 0.52$ fm \cite{Lewin:1995rx}.

The recoil spectrum, that is, the differential scattering rate per unit target mass for a \dm particle (whether a WIMP, NDM state, or otherwise) to scatter off a SM nucleus, is given by
\beq
\frac{dR}{dE_R} = \frac{1}{m_N}\int_{v>v_{\text{min}}}d^3 \mathbf{v} \; \frac{\rho_X}{m_X} f(\mathbf{v})v \frac{d\si_{XN}}{dE_R}\bigg|_v,
\eeq 
where $v_{\text{min}} = \sqrt{\frac{E_R m_N}{2\mu_{XN}^2}}$ (for non-relativistic scattering) is the minimum velocity required to obtain a recoil energy $E_R$, $\mu_{XN}$ is the reduced mass of the SM nuclei-\dm state system, $\si_{XN}$ is the SM nuclei-\dm scattering cross-section, $\rho_X$ is the local \dm density, $m_N$  and $m_X$ are the masses of the SM-nucleus and \dm state respectively, and $f(\mathbf{v})$ is the \dm velocity distribution.

Now specifying to our case of a NDM state with $k$ dark nucleons we follow \cite{Hardy:2014mqa} and assume that the underlying scattering process between the SM and DNs is isotropic and velocity-independent. This allows us to write the differential event rate as
\beq\label{eq:RecoilSpectrum}
\begin{split}
\frac{dR}{dE_R} &= 
g(v_{\text{min}}) \frac{\rho_k}{2\mu_{kn}^2 m_1}A^2 k \si_0 F_N(q)^2 F_k(q)^2,
\end{split}
\eeq
where $\si_0$ is defined as the zero-momentum transfer DN-SM nucleon scattering cross-section (and so $F_k(0) = F_N(0) = 1$), $\mu_{kn}$ is the reduced mass of the DN-SM nucleon system, $m_1$ is the mass of a single DN, $q = \sqrt{2m_N E_R}$, and
\beq 
g(v_{\text{min}}) = \int_{v>v_{\text{min}}}\hspace{-7mm}d^3\mathbf{v}\frac{f(\mathbf{v})}{v}.
\eeq

Note that the event rate is proportional to a linear factor of $k$ due to the cancellation of one power of the coherence factor ($k^2$) with a power of $k$ in the denominator coming from the mass of the dark nuclear state, $m_k=km_1$. As we will only consider the case of a single size of dark nuclei, the density of \kdn states, $\rho_k$, is the total local \dm density.

In order to determine the recoil spectrum reconstructed by an experiment, one must take into account the detector's efficiency $\ep_\text{eff}(E_R)$, and energy resolution $\si(E_R)$. The reconstructed recoil spectra are then found using
\beq\label{eq:RRS}
\begin{split}
\frac{dR_\text{rec}}{dE_R} &= \int dE_R' \ep_\text{eff}(E_R') \ka(E_R,E_R') \frac{dR}{dE_R'} \\&\approx \int dE_R' \frac{\ep_\text{eff}(E_R')}{\sqrt{2\pi}\si(E_R')}\exp\left(-\frac{(E_R-E_R')^2}{2\si^2(E_R')}\right)\frac{dR}{dE_R'}
\end{split}
\eeq
where $\ka(E_R,E_R')$ is the energy response function, which we have taken to be a Gaussian.

The predicted number of events seen by a detector is then
\beq
N = \left(\int_{E_\text{low}^{'}}^{E_\text{up}^{'}} dE_R \frac{dR_\text{rec}}{dE_R}\right)\times \text{Exposure},
\eeq
where $E_\text{low}^{'}$ and $E_\text{up}^{'}$ are the limits of the reconstructed energy window used in the analysis, and the exposure is the fiducial mass of the detector multiplied by its livetime.

\section{Nuclear Dark Matter in Tonne-Scale Direct Detection Experiments}\label{sec:Sens}

\subsection{Energy Resolutions, Efficiencies and Energy Windows.}\label{sec:DEAP}
In this section we detail the parameterisation of the predicted energy resolution, detector efficiency and recoil energy window used in our analysis for each experiment we consider. The recoil energy spectrum, \Fref{eq:DNFormFactor}, has two features that play an important role in the analysis. The first is the oscillatory nature of the Bessel function leading to non-decreasing sections of the recoil spectrum. In order to identify these features the experiment needs to have sufficiently good energy resolution. For this reason we have included an analysis with \deap, which has a particular focus on maximising the detected yield per unit of deposited energy (photoelectrons per keVee) and electronics designed to give excellent resolution over orders of magnitude in dynamic range, thereby producing a leading energy resolution. 

The second important feature of the NDM recoil spectrum is that the first few peaks at low energy dominate the event rate. This means that a low recoil energy threshold is an advantage for maximising the event rate. In addition, for large $k$ states the frequency of oscillations is higher and so in order to distinguish between NDM and WIMP models we need to detect as many events as possible in the non-decreasing part of the spectrum, motivating the lowest possible recoil energy threshold. For this reason, we have included an analysis with \xenon, which combines a low recoil energy threshold with tonne-scale fiducial mass. 

The size of the full signal energy window and the efficiency within the window for each experiment also plays a role in determining how many of the NDM form factor oscillations are in principle observable in each experiment. We start with the details of \deap.

\deap is a single-phase scintillation detector with a 1000 kg fiducial mass target of liquid argon, which will begin its search in the near future and results are anticipated over the next few years. In the single-phase detector, an array of photomultiplier tubes (PMTs) views the target over the full solid angle of 4$\pi$ to detect the scintillation light of any possible \dm scatterings. This approach maximises the number of photoelectrons (PE) detected, which is a function of the recoil energy (in keV) imparted by the collisions. To predict how WIMPs and NDM will be reconstructed we ran Monte Carlo simulations of $\sim 8\times 10^6$ events in a DEAP-like detector simulation, with parameters drawn from \cite{Amaudruz:2014nsa}. The energy resolution and efficiency were then found by fitting functions to the resulting distributions of measured PE versus generated recoil energy resulting in 
\beq
\begin{gathered}\label{eq:derecon}
\si(E_R) = 0.54 \,\kev \sqrt{E_R/\kev} - 0.048\,\kev (E_R/\kev) + 0.0026\,\kev(E_R/\kev)^{3/2}, \\ \ep(E_R) = 0.56 \,\erf(0.044\;E_R/\kev).
\end{gathered}
\eeq
We consider events in a recoil energy range of approximately $20<E_R<120$ keV (in fact selecting events with $50 < E_R < 300$ detected PE), since below this there are not enough PEs to accurately reconstruct the position of the event, and above there are too few \dm scatters.

\xenon is a two-phase scintillation detector with 1000 kg fiducial mass of liquid xenon target. \xenon is currently in the commissioning phase and will begin its physics run shortly. The relevant information for \xenon was taken from \cite{Aprile:2015uzo}, with an approximation for the
energy resolution given by \cite{MarcoSelvi}. This and the efficiency were then taken to be
\beq
\begin{gathered}\label{eq:xerecon}
\si(E_R) = -0.47 \,\kev + 0.65\,\kev \sqrt{E_R/\kev} + 0.02\,\kev(E_R/\kev), \\ \ep(E_R) = 0.4,
\end{gathered}
\eeq
where the energy window $4 < E_R < 50 \,\kev$ is used.

We also calculate the bound on NDM from the existing published LUX results \cite{Akerib:2015rjg}. All requisite information for LUX, apart from the energy resolution,  was taken from a recent improved analysis of the 2013 data \cite{Akerib:2015rjg} and the first results report \cite{Akerib:2013tjd}. For the energy resolution, we use the resolution for electron recoils from a more recent calibration of LUX using tritium decays \cite{Akerib:2015wdi}, $\si(E_R) = 0.32 \,\kev\sqrt{E_R/\kev}$. This was then converted into the energy resolution for nuclear recoils using a Lindhard model quenching factor \cite{Lindhard} with a Lindhard factor of $0.174$, as given in \cite{Akerib:2015rjg}. LUX's efficiency is taken directly from Figure 1 in \cite{Akerib:2015rjg}, within an energy window $1.1 < E_R < 50 \,\kev$.

\subsection{Example Recoil Spectra at \deap and \xenon}\label{sec:example}
\begin{figure}[ht!]
\begin{center}
\begin{subfigure}[b]{0.49\textwidth}
\includegraphics[width=\linewidth]{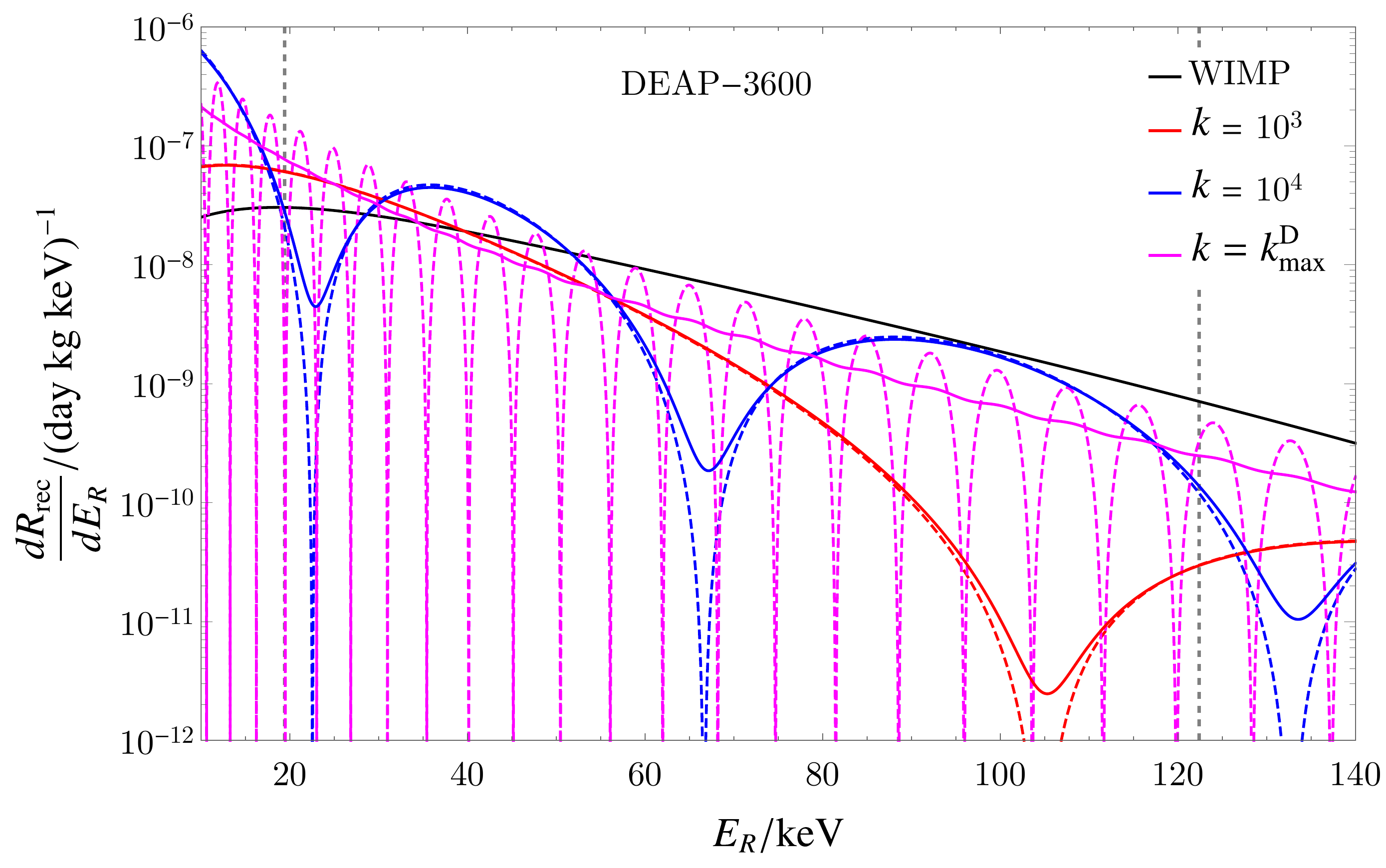}
\caption{}
\label{fig:RRSDEAP}
\end{subfigure}
\begin{subfigure}[b]{0.49\textwidth}
\includegraphics[width=\linewidth]{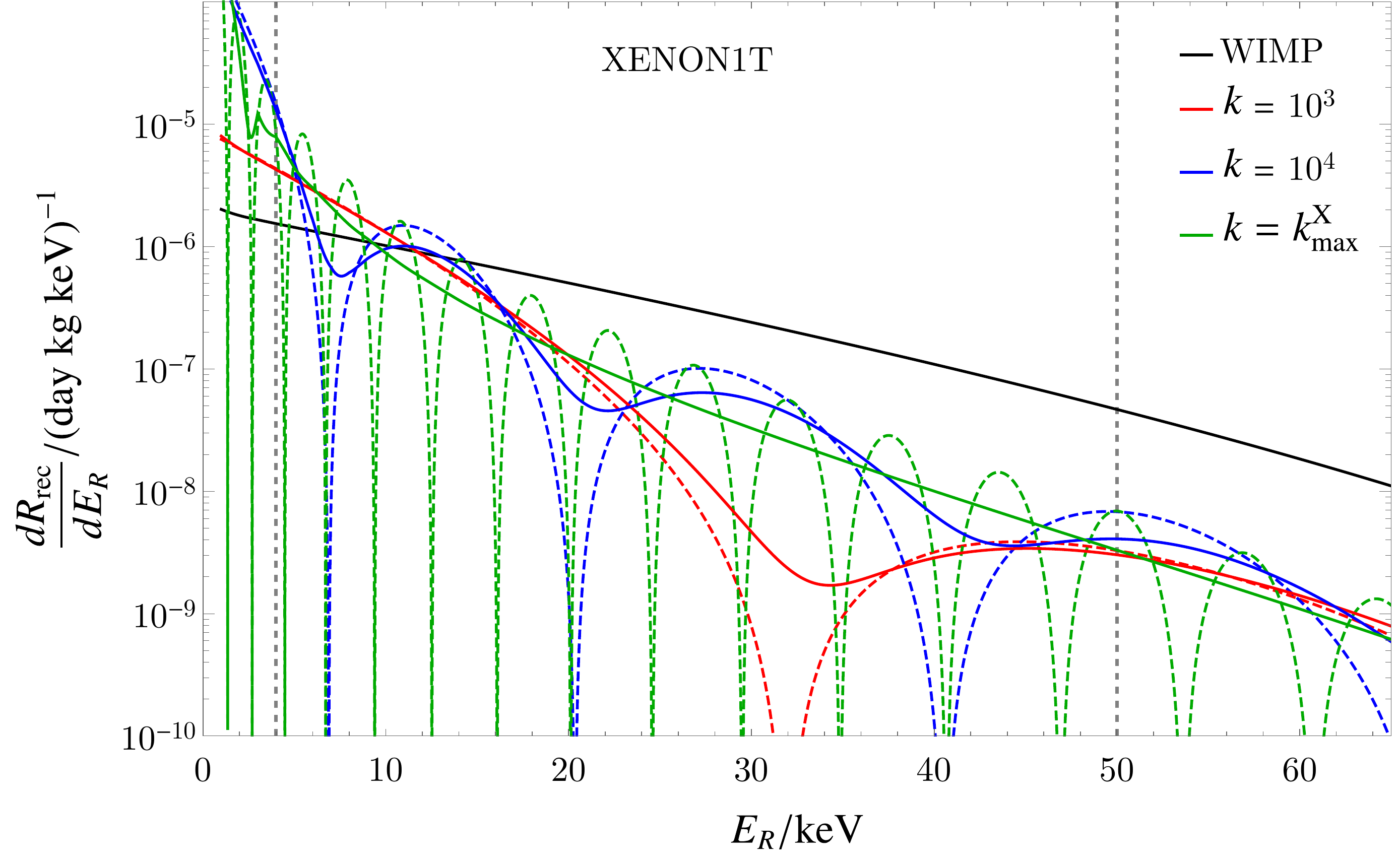}
\caption{}
\label{fig:RRSxenon}
\end{subfigure}
\caption{(a): Comparison of the recoil energy spectra for NDM along with that of a 100 GeV WIMP for \deap. Spectra are plotted for three different $k$ values: $10^3$ (red), $10^4$ (blue), and $k^{\text{D}}_{\text{max}}=6.5\times10^6$ (purple). (b): Comparison of the recoil energy spectra for NDM along with that of a 100 GeV WIMP for \xenon. Spectra are plotted for three different $k$ values: $10^3$ (red), $10^4$ (blue), and $k^{\text{X}}_{\text{max}}=5.8\times10^5$ (green). The bold and dashed lines represent the spectra with and without the finite energy resolution and efficiencies taken into account for each experiment using the experimental parameters in \Fref{eq:derecon} and \Fref{eq:xerecon} for \deap and \xenon respectively. In the NDM cases we have set $R_1 = 1$ fm and $m_1 = 1$ GeV. Vertical dashed lines represent the limits of the energy windows of the two experiments. The WIMP-SM nucleon scattering cross-section was fixed to be $10^{-46}\text{ cm}^2$ and the NDM cross-sections were scaled such that the integrated rates were equal to that of the WIMP across the energy window of each experiment.}
\label{fig:RRS}
\end{center}
\end{figure}

To illustrate the phenomenology of NDM scattering, we compare the reconstructed recoil spectra of NDM with that of a WIMP in \deap and \xenon. In \Fref{fig:RRS}, the recoil energy spectra for a range of $k$ values are shown for each experiment. For each value of $k$ we display the spectra with and without (bold and dashed lines respectively) the inclusion of the detector energy resolution and efficiency as detailed in \Fref{sec:DEAP}. We have fixed the NDM model parameters $R_1$ and $m_1$ to 1 fm and 1 GeV respectively in analogy with the SM. We have no fundamental guide for what these parameters should be in the dark sector, so we have chosen to take the SM values as a place to start. These parameters will remain fixed at these values throughout the rest of this paper. Also shown in each plot of \Fref{fig:RRS} is the reconstructed recoil spectrum of a standard WIMP with a mass and elastic scattering cross-section per SM nucleon of 100 GeV and $10^{-46}\text{ cm}^2$ respectively. We use a Maxwell-Boltzmann \dm velocity distribution for the WIMP spectrum and assume the scattering is isotropic and momentum-independent. 

In most cases even with detector effects included the oscillations of the \kdn form factor can be seen, which contrasts with the monotonically decreasing WIMP spectrum\footnote{The WIMP spectrum will start increasing again due to the oscillations in the SM form factor, but this occurs far beyond the maximum energy window of \deap and \xenon that we consider here.}. The period of the oscillations in the NDM recoil energy spectrum decreases with $k$, as the size of the \kdn states increase. It is worth noting at this point that the slope enveloping these oscillations also steepens with $k$, due to the inverse dependence of the form factor on $k$ as shown in \Fref{eq:DNFormFactor}.

Comparison of the solid and dashed curves in \Fref{fig:RRS} demonstrates the effect of the finite experimental energy resolution. The sharpness of the features are smoothed out leading to less significant troughs in the spectrum and simultaneously decreasing the height of the peaks. As $k$ increases the period of the oscillations will decrease and the effect of the energy resolution smearing becomes more prominent. As the oscillation period approaches the energy resolution of the detector it is clear that there is a value $k_\text{max}$ beyond which the characteristic oscillations of the form factor can no longer be resolved. In particular, the non-decreasing parts of the spectrum are smoothed away leaving a monotonically decreasing spectrum. This transition is important for our analysis as it represents the value of $k$ beyond which the distinguishability between the NDM spectrum and the WIMP spectrum is no longer determined by the non-decreasing sections of the spectrum but rather by the shape of the slope as a whole. We note that a non-standard velocity distribution may allow a WIMP spectrum to mimic the NDM spectrum for dark nuclei with $k>k_\text{max}$, while for lower $k$ values the non-decreasing sections of the NDM spectrum would still allow an experiment to distinguish NDM from a WIMP spectrum no-matter what the velocity distribution. 

We can determine $k_\text{max}$ by finding when the period of the oscillations at the threshold energy\footnote{We compare at the threshold energy as this is where the dominant part of the signal will come from and is where the period of the oscillation is smallest.} is approximately equal to the width of the response function, $\si(E^{\text{th}}_R)$, where $E^{\text{th}}_R$ is the threshold energy of the detector. This leads to
\beq\label{eq:kup}
k_\text{max} \sim \left(\frac{\pi^2}{4 m_N R_1^2 \left(E^{\text{th}}_R - \sqrt{(E^{\text{th}}_R)^2 - \si(E^{\text{th}}_R)^2}\right)}\right)^{3/2}.
\eeq
Beyond $k_\text{max}$ we expect the dominant part of the NDM spectrum to be effectively a monotonically decreasing function of the recoil energy. In \Fref{fig:RRS}, we have plotted the $k_\text{max}$ for each detector, $k^{\text{D}}_\text{max}$ for \deap and $k^{\text{X}}_\text{max}$ for \xenon, and in each case we can see that the spectra including the detector effects have been smoothed out leaving no rising sections.

Conversely if we decrease the value of $k$ sufficiently we will eventually generate a recoil spectrum that has no rising sections within the energy window of the experiment. The NDM form factor in this region only produces a modification in the slope of the recoil energy spectrum, which again could be mimicked easily by a WIMP spectrum. Defining $k_\text{min}$ as the DN number at which the first trough in the recoil spectrum enters the energy window, we find
\beq\label{eq:klow}
k_\text{min} \sim \left(\frac{4.5}{R_1 \sqrt{2 m_N E_\text{up}}}\right)^3,
\eeq
where $E_\text{up}$ is the upper end of the sensitivity energy window for a given experiment. With these definitions, for each experiment there is a range of $k$ values in which the distinctive rising features of the dark form factor may be seen within the energy window of a detector; outside this region we expect NDM spectra to be monotonically decreasing.

For \deap this $k$ range is
\beq\label{eq:krange}
800 \lesssim k\left(\frac{R_1}{1 \text{fm}}\right)^3 \lesssim 6.5\times 10^6.
\eeq
and for \xenon the equivalent range is
\beq\label{eq:krangeX1T}
500 \lesssim k\left(\frac{R_1}{1 \text{fm}}\right)^3  \lesssim 5.8\times 10^5.
\eeq
The lower bound in \xenon is slightly lower than in \deap as the first dip appears within the energy window sooner due to the larger mass of the xenon nucleus, leading to a shorter period in the form factor oscillation. Similarly, comparing \deap and \xenon for $k=10^3$ and $k=10^4$ we note that for the same value of $k$ the period of oscillations in the spectra is shorter for \xenon again due to the larger mass of the xenon nucleus producing a larger value for the momentum transfer, $q$, for a given recoil energy. In addition for the majority of the relevant range of recoil energies, \deap has a better energy resolution potentially allowing it to distinguish (using rising sections of the spectra) dark nuclei with larger $k$ values compared to \xenon as the transition to a monotonically decreasing spectrum will occur at larger values of $k$ for \deap.  

In \Fref{sec:Likelihood} we perform an analysis to determine the number of signal events required to distinguish the NDM spectrum from that of a WIMP. For $k$ values within the ranges in \Fref{eq:krange} and \Fref{eq:krangeX1T} the number of signal events required to identify the NDM spectrum will depend on two main properties; where the first (and sometimes subsequent) rising section of the spectrum appears in the energy window and what the amplitudes of the oscillations are after detector energy resolution has been taken into account.

\subsection{Current and Projected Exclusion Limits}\label{sec:Limits}
\begin{figure}
\centering
\includegraphics[width=0.75\linewidth]{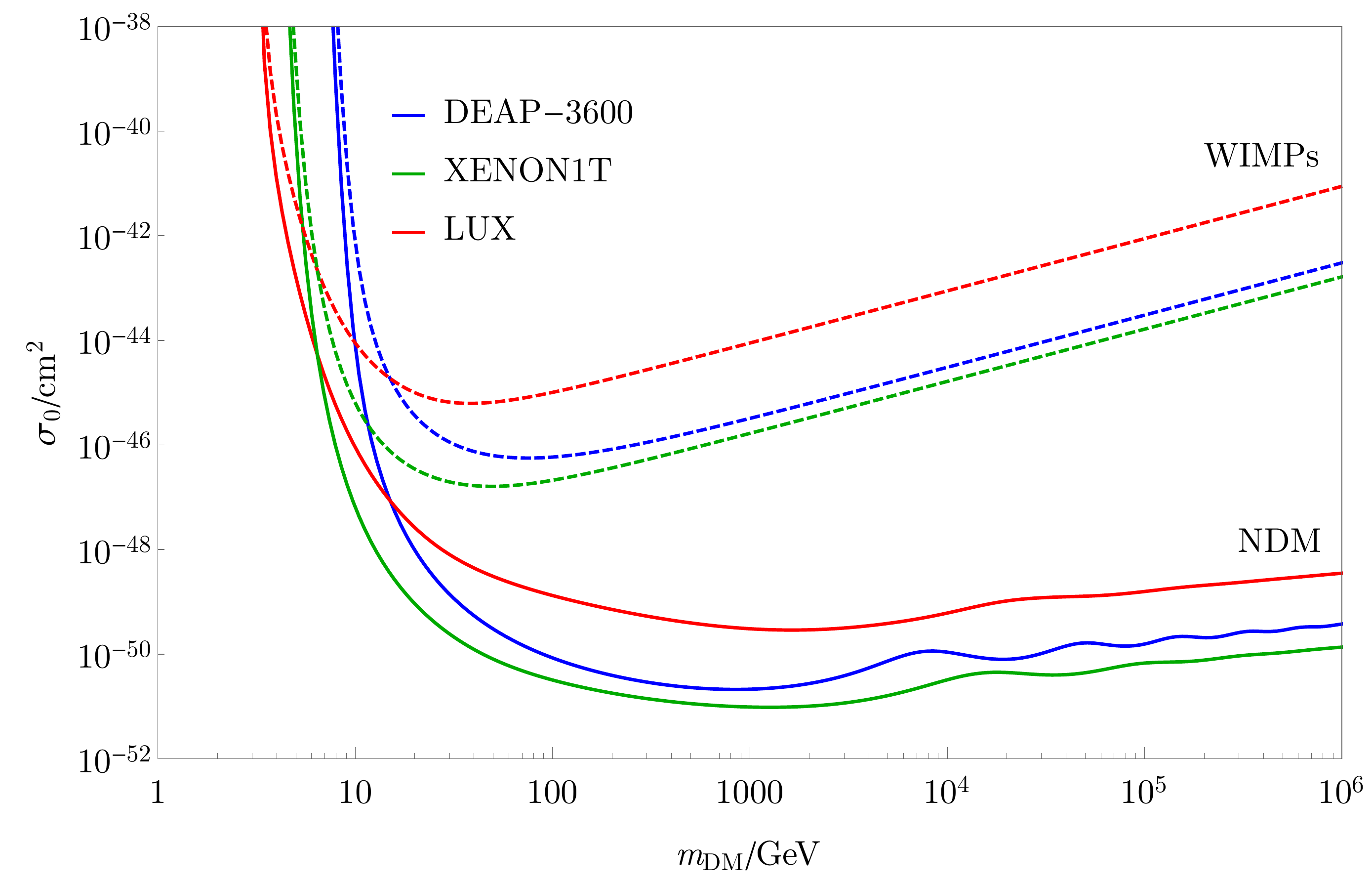}
\caption{Projected limits at 90\% CL for $\si_0$, the DN-SM nucleon scattering cross-section (in bold), against the \dm mass for \deap (blue) \cite{Amaudruz:2014nsa} and \xenon (green) \cite{Akerib:2015rjg}, and compared with the strongest current limits from LUX (red) \cite{Aprile:2015uzo}. Also shown are the WIMP limits in the same detectors (dashed lines), where $\si_0$, is interpreted as the WIMP-SM nucleon scattering cross-section. For the NDM limits we have taken $m_1 = 1$ GeV and $R_1 = 1$ fm.}
\label{fig:Limits}
\end{figure}

While our main focus is to examine whether we can distinguish NDM from WIMP signal events, we finish this section with a discussion of the current and projected limits on models of NDM if neither \deap or \xenon observe any candidate signals. \Fref{fig:Limits} shows the 90\% confidence level (CL) limits we calculate on $\si_0$, the DN-SM nucleon scattering cross-section for NDM, compared with the equivalent limits for a standard WIMP, each as a function of the mass of the \dm state. For NDM, the mass of the \dm state is given as $m_k=k\;m_1$, where we have fixed the mass of an individual DN to $m_1 = 1$ GeV. As a result the x-axis is equivalent to the number of DNs, $k$, for the nuclear states. 

In both the NDM and WIMP cases, we show the projected limits for \deap and \xenon and the current limit derived from LUX. All NDM limits are calculated using \Fref{eq:RRS} with the requisite detector information from \Fref{sec:DEAP}. Our calculated WIMP limits (dashed lines in \Fref{fig:Limits}) are in good agreement with the results presented in the experimental papers \cite{Amaudruz:2014nsa,Akerib:2015rjg,Aprile:2015uzo}.

Comparing the NDM limits with those for WIMPs, from \Fref{eq:RecoilSpectrum} we see that the constraint on the DN-SM nucleon cross-section is much stronger than for the WIMP. In the lower mass region below 100 GeV, the ratio of the NDM and WIMP limits for a given \dm mass goes approximately as $k^2$ as in this region the dark form factor is still close to unity. Towards higher masses the dark form factor suppresses the overall rate and therefore weakens the limits. The limiting behaviour of the event rate for large $k$ goes as $\sim k^{-1/3}$ and as a result the elastic scattering cross-section limit for dark nuclei increases as $k^{1/3}$ in contrast to the usual WIMP limit, which increases linearly with $m_{\text{DM}}$.

In the high mass region the NDM limits also exhibit oscillations (due to the dark form factor), which are more pronounced in the case of \deap. These arise because as $k$ (and therefore $m_k$) increases there are $k$ values for which the experiment energy window contains more peaks than troughs and visa-versa, which results in increased (or decreased) sensitivity. This effect is less pronounced in LUX and \xenon as the energy resolution is not as good in the high recoil energy region; this means the amplitude of the oscillations in the recoil spectra is not reconstructed as sharply, which results in the flatter limit shape. The scale of the NDM limit largely depends on the overall count rate above the energy threshold \ie if \deap's lower energy threshold could be reduced by a factor of $\sim 2$, below $\sim 8$ keV, then it could be more sensitive to NDM in the high mass region than \xenon.

The overall scale of the limits on the DN-SM nucleon scattering cross-section is several orders of magnitude more restrictive than that for the WIMP case. In the example of a scalar interaction between a SM quark and a DN (scalar or fermion), we can translate the limits from direct detection to an upper limit on the DN annihilation to SM quarks. We find a result that is many orders of magnitude below what is required for achieving the correct abundance via standard freeze-out. This is no great surprise as it has been known for some time \cite{MarchRussell:2012hi} that it is very difficult to have a sufficiently large \dm annihilation rate to SM quarks for asymmetric freeze-out to produce an abundance that is determined by the \dm asymmetry. The limits from direct detection and mono-jet searches at the LHC rule out the viable parameter regions involving interactions with quarks or gluons. The process of asymmetric freeze-out in NDM models can proceed, however, via the annihilation into lighter hidden sector states \cite{Hardy:2014mqa}. These hidden sector states may also be limited in a model-dependent way by astrophysical constraints, the details of which are beyond the scope of this work.

\section{Identifying Nuclear Dark Matter}\label{sec:Likelihood}
Perhaps the most interesting question is: if \deap or \xenon do see a candidate signal above the background, how many events are required to identify nuclear dark matter? In this section we calculate the number of events required to distinguish between the NDM and WIMP hypotheses (\nreq) at a given CL.

In the absence of knowledge about the nature of dark matter particles, we compare each NDM spectrum (of varying $k$ values) with the most similar WIMP-induced recoil energy distribution. This most similar WIMP mass can be different for each $k$, and therefore the number of events required to effectively distinguish between the hypotheses will be different in each case. 

We start by finding the WIMP spectrum which is the most indistinguishable, \ie requires the highest number of events to distinguish, from each hypothesised NDM spectrum. To this end we define a test statistic, $\lambda$, as a negative log likelihood (NLL) ratio;
\begin{equation} \label{eq:teststat}
\lambda = -2\ln \left( \frac{\mathcal{L}_{\rm{NDM}}}{\mathcal{L}_{\rm{WIMP}}} \right),
\end{equation}
where $\mathcal{L}_{\rm{NDM/WIMP}}$ is the likelihood of an event
occurring due to a NDM or WIMP \dm candidate. Here values
greater than zero correspond to data sets which are more WIMP-like
than NDM-like. An unbinned extended likelihood is used giving
\begin{equation} \label{eq:likelihood}
\mathcal{L}_{\rm{NDM/WIMP}} = {\rm Poisson}\left\{N_{\rm{obs}},N_{\rm{exp}} (\sigma_0,k,m_1) \right\} \prod_{i=1}^{N_{\rm{obs}}} f(E_i | k,m_1).
\end{equation}
where $f(E_i | k,m_1)$ is the reconstructed recoil energy probability density function (PDF) for the WIMP ($k=1$) or NDM case. The mass $m_1$ (the mass of a single DN or WIMP, depending on the case), and elastic scattering cross-section $\sigma_0$, are allowed to take independent values for the NDM and WIMP cases. The scattering cross-section acts as an overall normalisation factor (under our assumptions of isotropic and velocity-independent scattering) in both distributions; as such, it only appears in the integral number of expected events $N_{\rm{exp}}$. The most indistinguishable case occurs when the expected number of events is the same for both NDM and WIMP dark matter, thus the Poisson term can be dropped from the likelihood ratio and only the shape of the distributions is compared.

For each hypothesised number of observed events the probability, $p$, of misidentifying NDM as WIMPs is determined from the PDF of $\lambda$, $g(\lambda | N_{\rm{obs}})$, where
\begin{equation} \label{eq:misprob}
p = \int_{0}^{\infty} g(\lambda | N_{\rm{obs}}) d\lambda.
\end{equation}
The PDF is built using a Monte Carlo simulation where events are created under the NDM hypothesis for a given $k$. A confidence level $C=1-p$ is then assigned which corresponds to the probability that a dataset will be correctly identified as NDM-like under the NDM hypothesis. 
\begin{figure}[ht!]
\begin{center}
\includegraphics[width=0.75\linewidth]{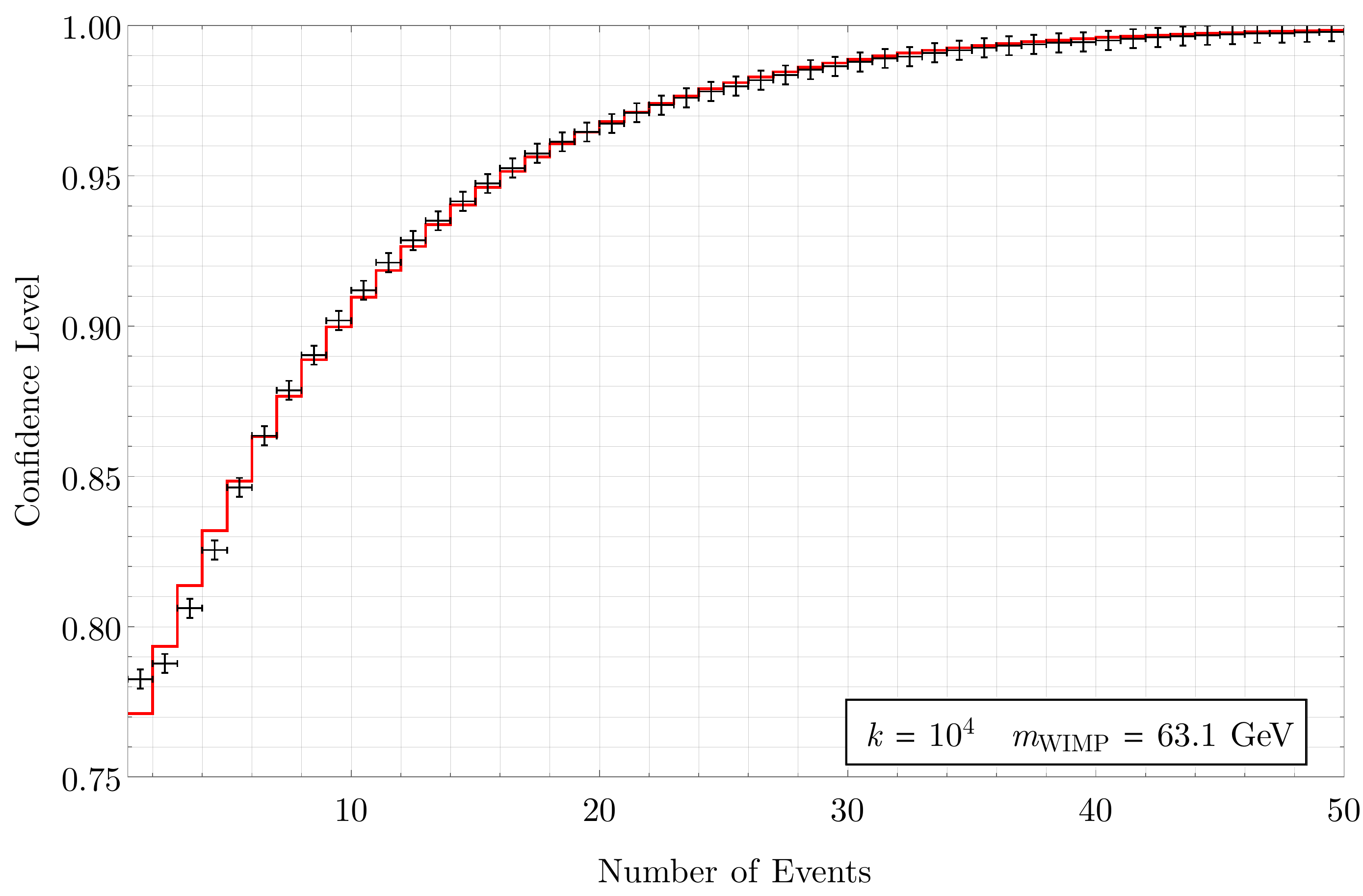}
\caption{The CL for each value of $N_{\rm{obs}}$ up to $N_{\rm{obs}} = 50$ for $k = 10^4$ ($m_1 = 1 \, \rm{GeV}$) compared to a $63.1 \, \rm{GeV}$ WIMP. The black crosses mark the values found from the Monte Carlo simulation, while the red line presents \Fref{eq:confidence} when fitted to this data. The integer number of events ($\lfloor{N_{\rm{obs}}}\rfloor$) is used in \Fref{eq:confidence}, which is reflected in the equation's step-like shape.}
\label{fig:coverage}
\end{center}
\end{figure}

The test was carried out for a range of $k$ values, logarithmically distributed between $k=10$ and $k = 10^{7}$, with $m_1 = 1$ GeV. This range accommodates those specified in \Fref{eq:krange} and \Fref{eq:krangeX1T}, where we expect the non-decreasing sections of the dark form factor to be visible within each detector's energy window. For the WIMP distributions a table was built comprising of 5000 WIMP mass spectra logarithmically distributed between $m = 10$ GeV and $m = 10^6$ GeV. This range accounted for all possible WIMP distribution shapes. 
When $k$ is small, we do not expect dark form factor effects to be significant, and therefore the most indistinguishable WIMP will likely have around the same mass as the \kdn state. This means WIMP spectra of masses below 10 GeV will not be relevant. Furthermore, the sensitivity of \deap and \xenon sharply falls below this mass (see \Fref{fig:Limits}), and so the potential for detection is greatly reduced. For WIMPs of mass $10^6$ GeV or greater, the spectra are indistinguishable (in shape) from each other, and so do not affect the analysis which is only shape-dependent.

The WIMP spectra table was scanned through for each value of $k$. The number of events required to distinguish between the NDM and WIMP hypotheses up to a 3 $\sigma$ CL was then determined for each WIMP mass. Generating adequate Monte-Carlo statistics to produce distributions of $g(\lambda | N_{\rm{obs}})$ for each of the masses and number of events is computationally expensive. Consequently, $10^5$ pseudo-experiments were used for all potential WIMP spectra and the distribution of CLs was determined by fitting a modified non-central chi-squared cumulative distribution function (CDF):
\begin{equation} \label{eq:confidence}
C(N_{\rm{obs}}) = 1 - Q_{\frac{a}{2}} \left( \sqrt{b},\sqrt{\frac{N_{\rm{obs}}}{c}} \right),
\end{equation}
where $Q_{\frac{a}{2}}$ is the Marcum Q-function. Here the parameters $a$, $b$, and $c$ are left as variables in the fit. 

This function fits well to all CL distributions, one of which is illustrated in \Fref{fig:coverage}. The CL is shown for each value of $N_{\rm{obs}}$ up to $N_{\rm{obs}} = 50$ for $k = 10^4$ ($m_1 = 1 \, \rm{GeV}$) compared to a $63.1 \, \rm{GeV}$ WIMP (the most indistinguishable). \Fref{eq:confidence} was fitted to these points with the inverse being used to determine the number of events required for each CL. Here excellent agreement is seen between the Monte Carlo data and the fit, supporting the use of \Fref{eq:confidence} in the analysis.
\begin{figure}[ht!]
\begin{center}
\begin{subfigure}[b]{0.49\textwidth}
\includegraphics[width=\linewidth]{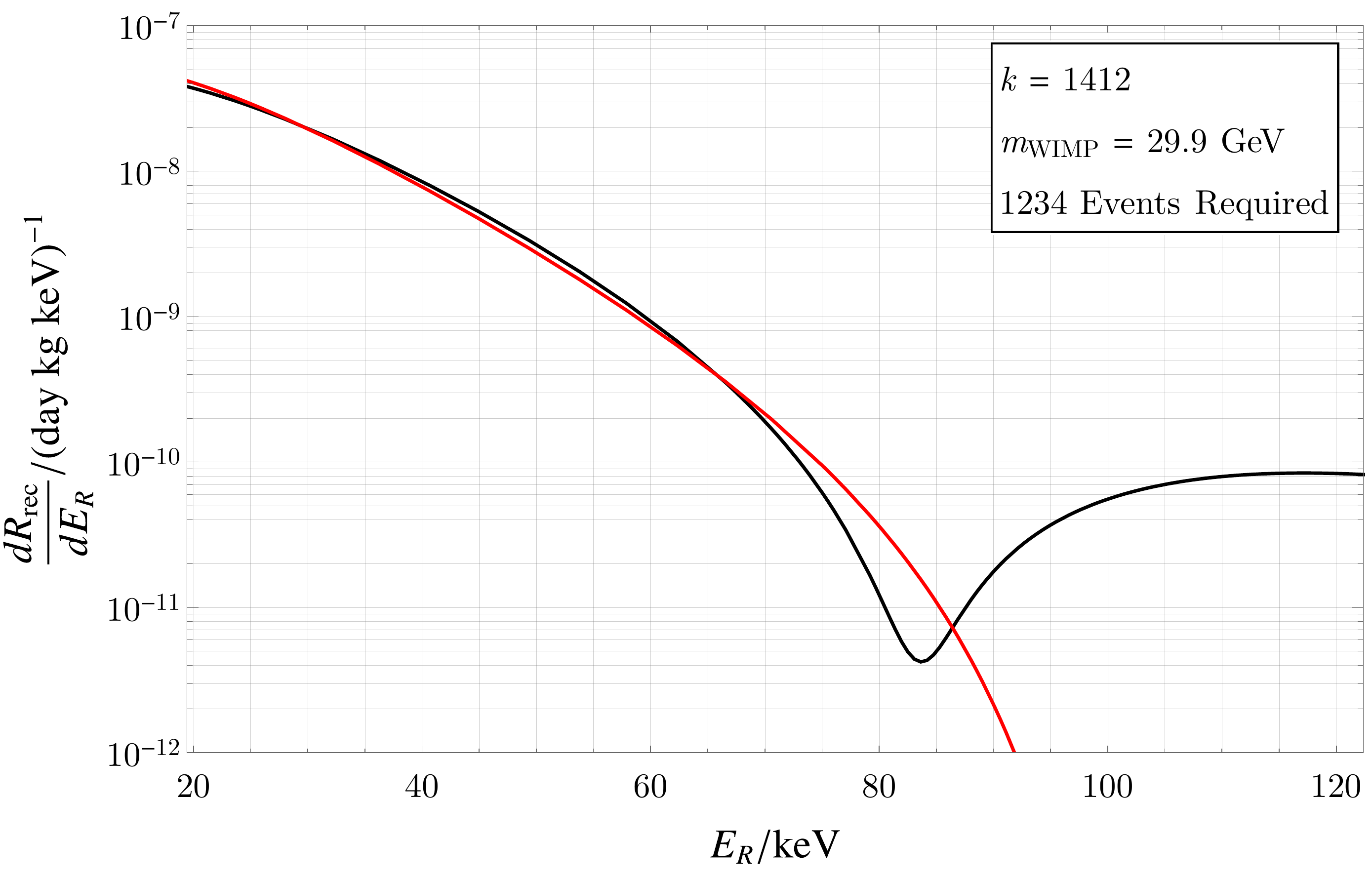}
\caption{}
\label{fig:k1412}
\end{subfigure}
\begin{subfigure}[b]{0.49\textwidth}
\includegraphics[width=\linewidth]{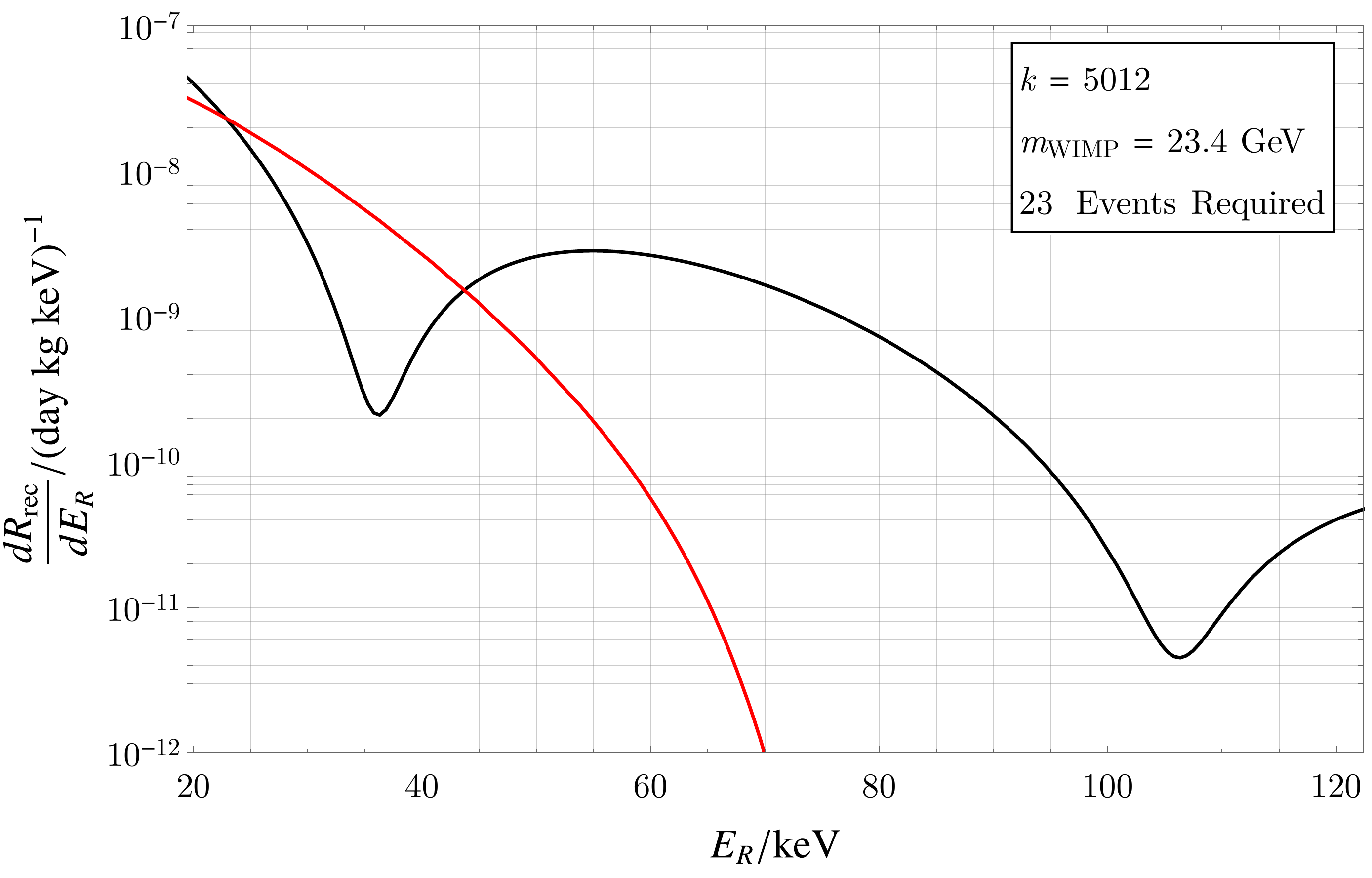}
\caption{}
\label{fig:k5012}
\end{subfigure}

\begin{subfigure}[b]{0.49\textwidth}
\includegraphics[width=\linewidth]{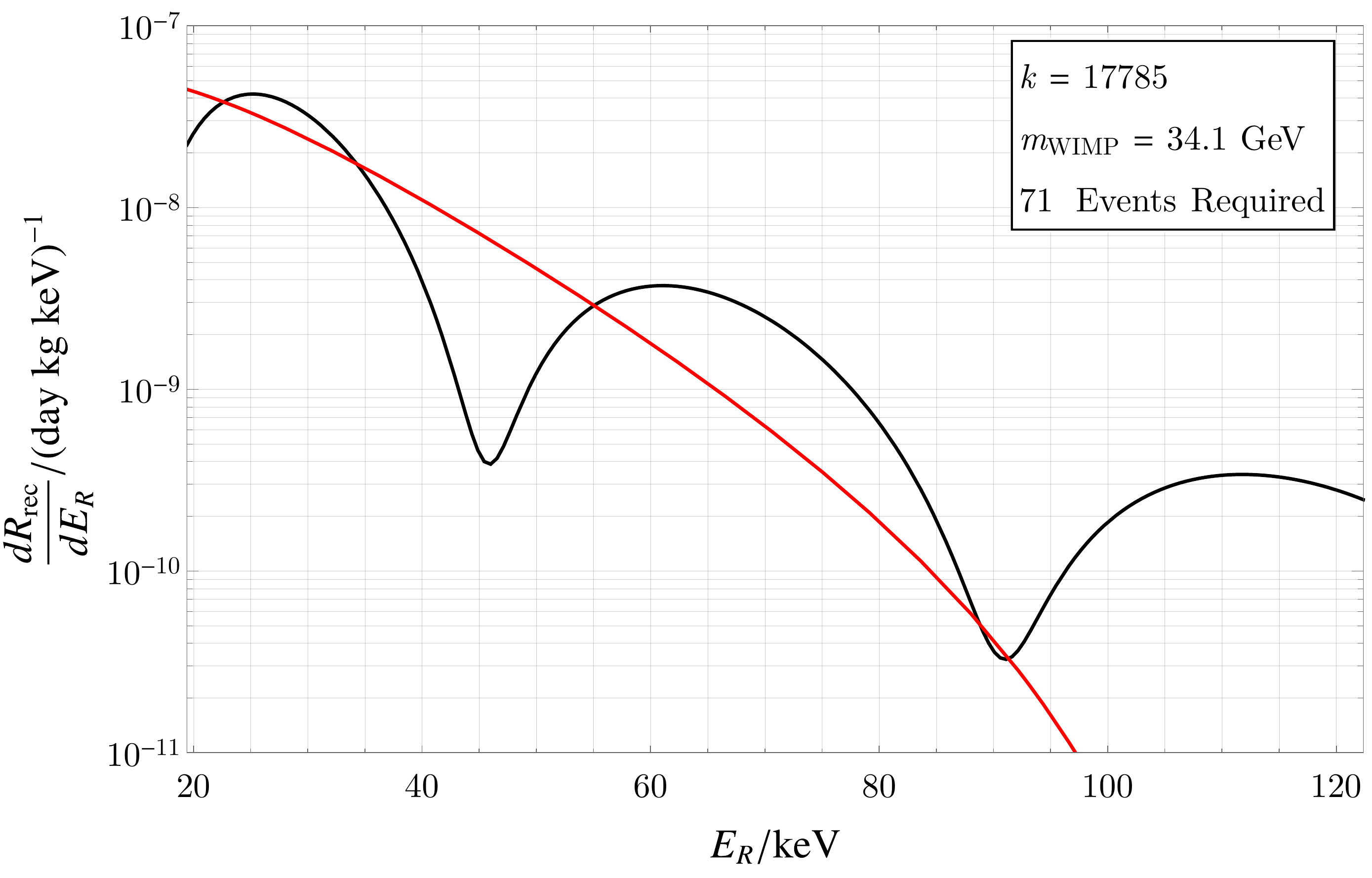}
\caption{}
\label{fig:k17785}
\end{subfigure}
\begin{subfigure}[b]{0.49\textwidth}
\includegraphics[width=\linewidth]{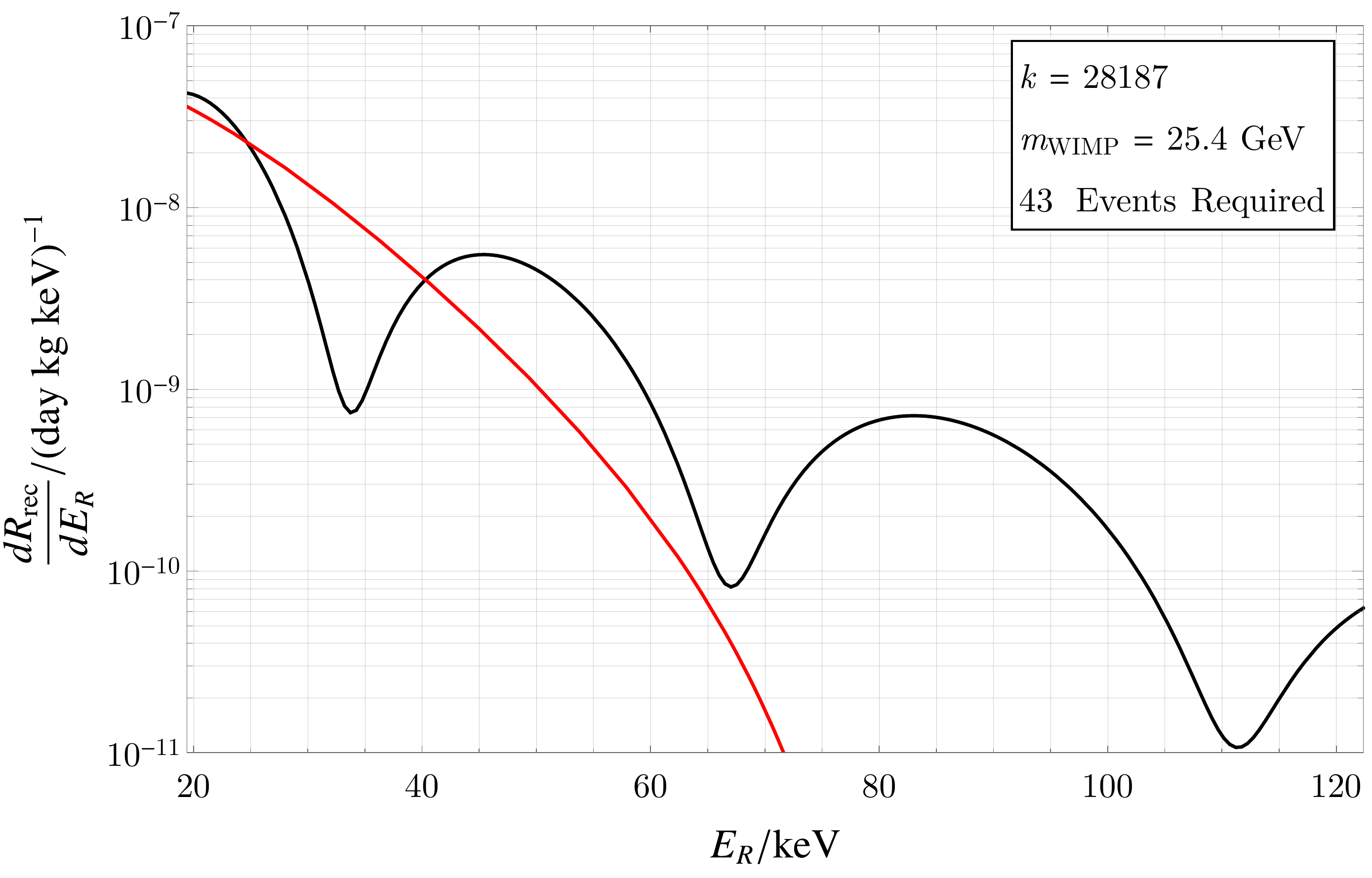}
\caption{}
\label{fig:k28187}
\end{subfigure}

\begin{subfigure}[b]{0.49\textwidth}
\includegraphics[width=\linewidth]{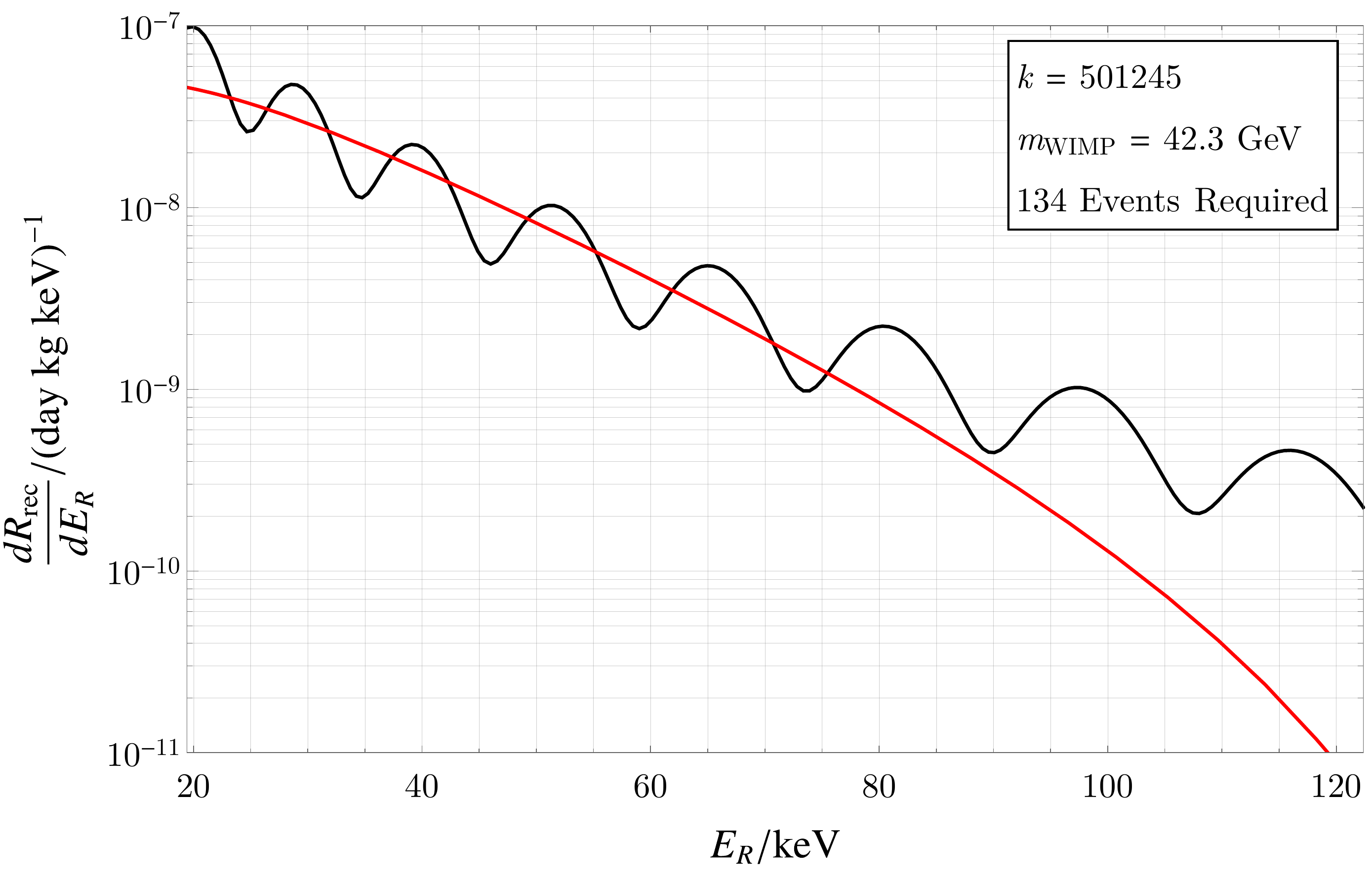}
\caption{}
\label{fig:k501245}
\end{subfigure}
\begin{subfigure}[b]{0.49\textwidth}
\includegraphics[width=\linewidth]{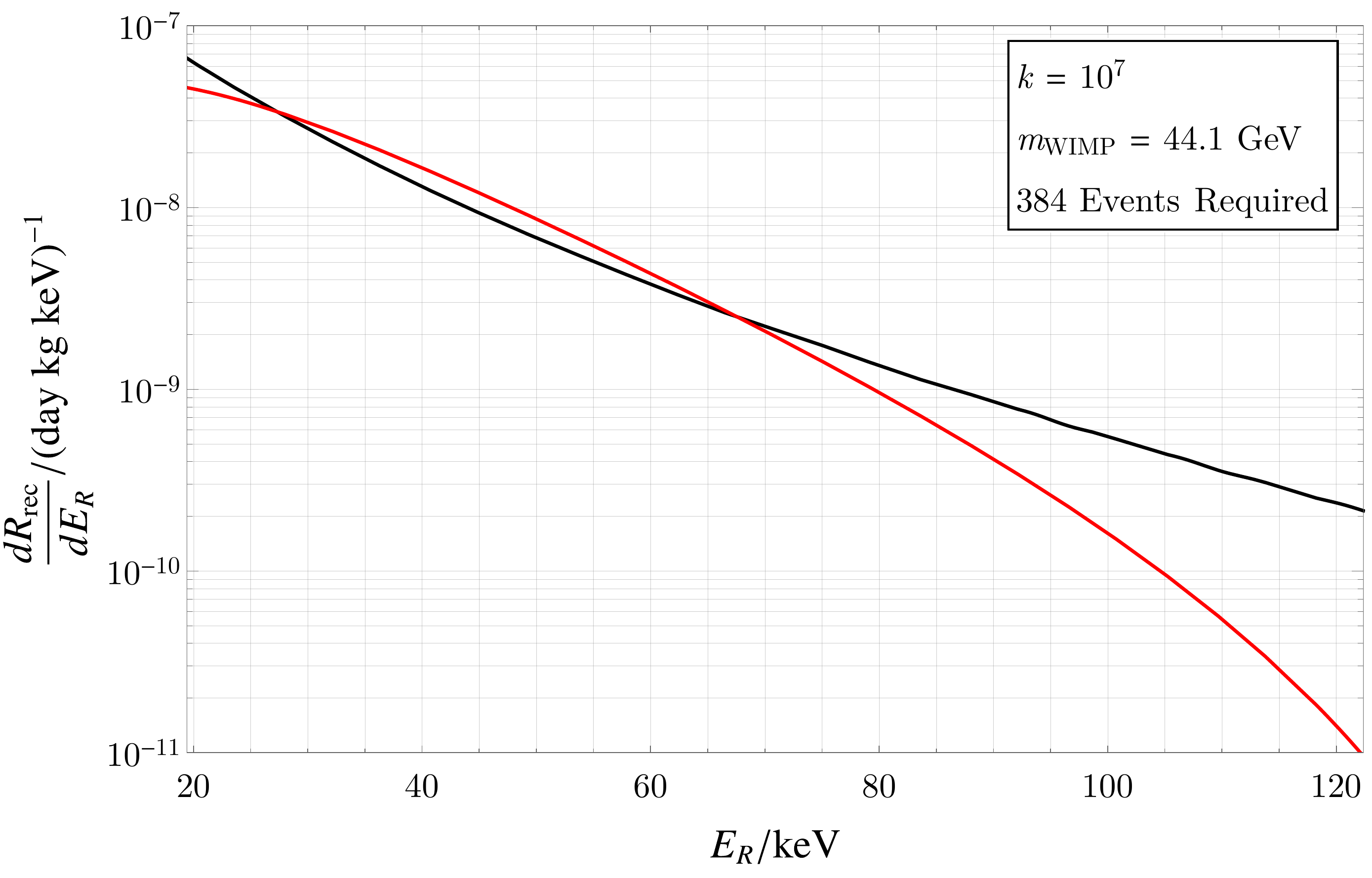}
\caption{}
\label{fig:k10pow7}
\end{subfigure}
\caption{Example NDM spectra (black) along with the most indistinguishable WIMP spectra (red), as reconstructed by \deap. Inset into each plot is the $k$ value, the mass of the best fitting WIMP ($m_\text{WIMP}$), and the number of events required to distinguish between the spectra to a $3\,\si$ CL.}
\label{fig:examplespectra}
\end{center}
\end{figure}
\begin{figure}[ht!]
\begin{center}
\begin{subfigure}[b]{\textwidth}
\centering
\includegraphics[width=0.75\linewidth]{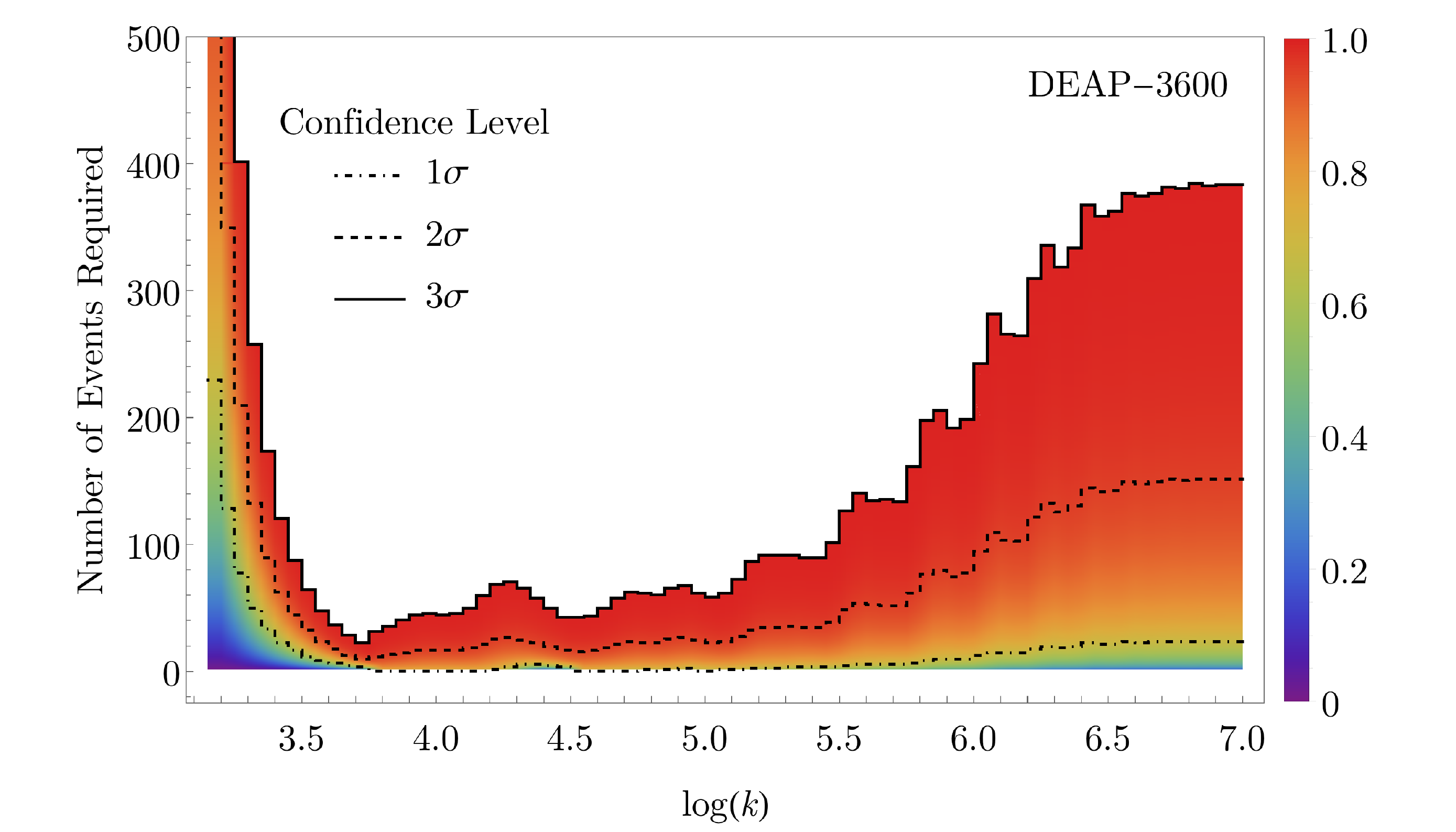}
\end{subfigure}

\begin{subfigure}[b]{\textwidth}
\centering
\includegraphics[width=0.75\linewidth]{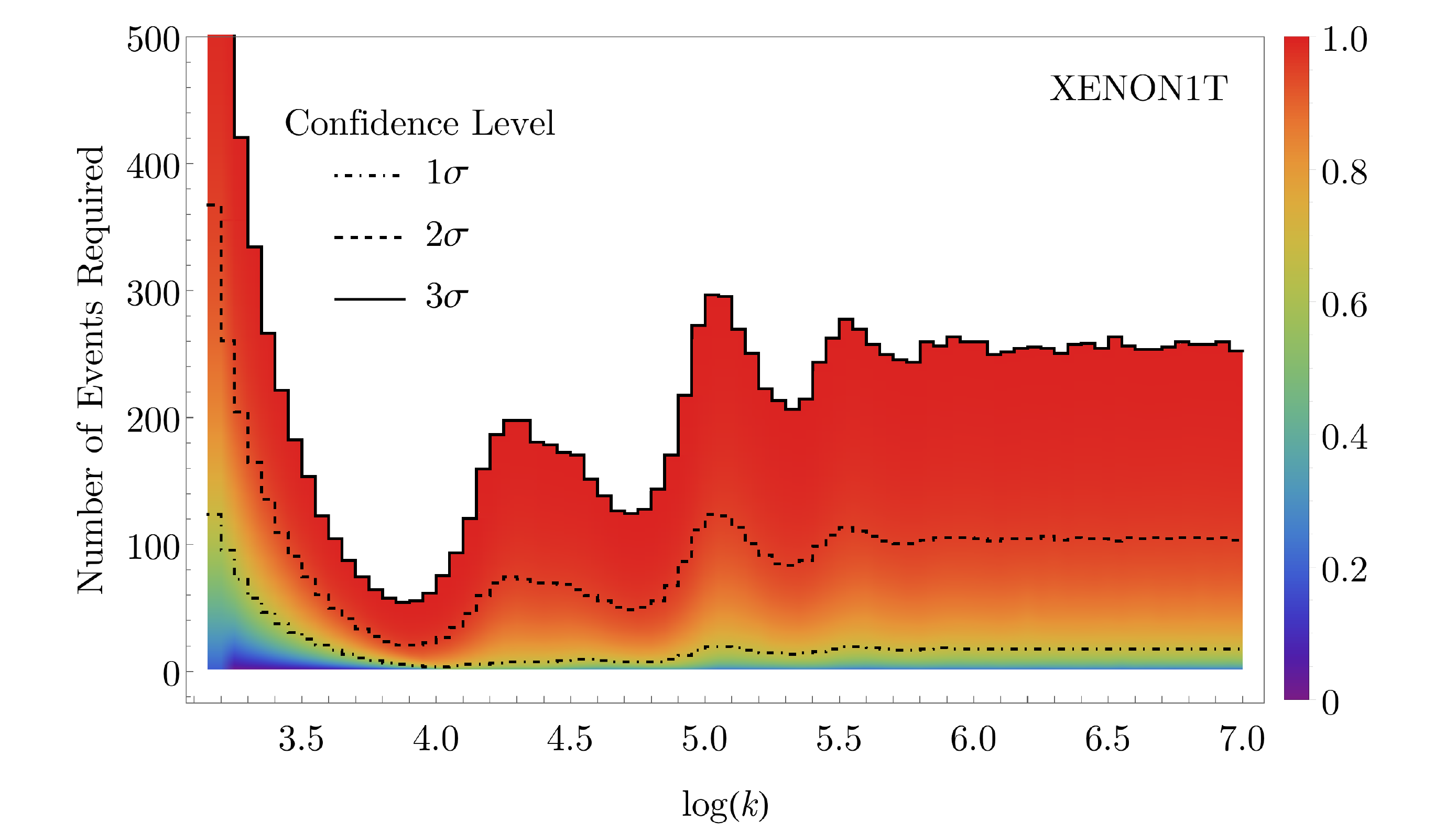}
\end{subfigure}
\caption{The number of events required to distinguish between values of $k$ and all WIMP masses to the stated CL in \deap (above) and \xenon (below). Here the CL given by \Fref{eq:confidence} is shown by the colour scale with 1, 2, \& $3\,\sigma$ shown as dot-dashed, dashed, and solid lines respectively. }
\label{fig:hclfit}
\end{center}
\end{figure}

\Fref{fig:examplespectra} shows a selection of NDM spectra as reconstructed by \deap, which are plotted across the energy window of the detector. In each sub-figure the spectrum is plotted alongside that of the most indistinguishable WIMP at that $k$ value. The top left panel (\Fref{fig:k1412}) shows the spectra for $k=1412$, which is the first bin in the $k$ range in \Fref{fig:hclfit}, where the number of events required to distinguish up to a $3\,\si$ CL is plotted as a function of $k$, for \deap (above) and \xenon (below). In the $k=1412$ case, the first trough in the recoil spectrum has just entered into the energy window, and the WIMP spectrum closely matches that of NDM. Subsequently the number of events required to distinguish to $3\,\si$ is large (1234). As $k$ drops below this value, \nreq gets larger and eventually the first trough leaves the energy window with distinguishability between WIMP and NDM spectra becoming effectively impossible. In the event of a signal in this case, one could only put an upper limit on $k$.

In \Fref{fig:k5012} (top right panel) where $k=5012$, the first trough occurs at a lower recoil energy, and the WIMP spectrum no longer closely follows that of NDM. Here, the relative contribution of the second peak is much greater than in \Fref{fig:k1412}, by around a factor of 100. Given that \nreq decreases by approximately this factor as well, this suggests the rise in the spectrum, which is irreproducible with a standard WIMP, acts as the dominant distinguishing feature of the spectra. This claim is further supported when we notice that \nreq in each of these cases is just enough to be able to resolve the second peak (and consequently, the first trough)\footnote{A crude criterion for a resolvable spectral feature is as follows: if the rate in the vicinity of the feature is a factor of $1/N$ smaller than the peak rate within the entire energy window, then at least $N$ events will be needed before events appear in this region, allowing it to be ``resolvable".}. Only 23 events are required in the case that $k=5012$, which corresponds to the lowest value across the entire range for \deap, \ie the deepest minimum in \Fref{fig:hclfit}.

This distinguishing mechanism is also at work in the case that $k=17785$ (seen in \Fref{fig:k17785}, the middle left panel), which corresponds to the first peak following the central minimum at $k=5012$ in \Fref{fig:hclfit}. Here the threshold is just at the beginning of a peak, which means the NDM spectrum follows the WIMP spectrum more closely than in the previous case and the second peak occurs later on, subsequently \nreq is larger. 

The next minimum in \nreq lies at $k=28187$, whose spectrum is seen in \Fref{fig:k28187} (middle right panel). Here the threshold intersects the oscillations approximately as it does in \Fref{fig:k5012} for $k=5012$, however more events are required to distinguish in the $k=28187$ case. This reflects the effect of the finite energy resolution which smears out the troughs, reducing their depth and making them less distinctive.

It is clear now that as $k$ is increased, \nreq oscillates as the spectra change between configurations such as those in \Fref{fig:k5012} \& \ref{fig:k28187}, and that in \Fref{fig:k17785}. These fluctuations are seen in \Fref{fig:hclfit} for both \deap and \xenon, the difference between them will be discussed later. The ability to distinguish between hypotheses for a particular value of $k$ in this region is dependent on the chosen energy window. If we were to change the lower threshold of the window, this would change the relative size of the second peak and therefore \nreq. The overall effect of this would be a phase change in the oscillations seen in \Fref{fig:hclfit}; the troughs and peaks would shift, but their amplitude would not change much. Changing the upper threshold by small amounts will have little effect, as the relative rate in this region is negligible (compared with the rate at threshold).

Inspecting \Fref{fig:k501245} (bottom left panel), where $k=501245$, which lies in the upwards trending region of \Fref{fig:hclfit}, we can see the period of the oscillations is much smaller, and the fluctuations have been smeared out significantly. With only $\mathcal{O}(10)$ events, several oscillations in the spectrum can now be probed (as opposed to only the first), however their amplitude has been reduced to such an extent that 134 events are required to distinguish to $3\,\si$. The upwards rise in \Fref{fig:hclfit} reflects the increase in smearing making distinction harder.

In the last case seen in \Fref{fig:k10pow7} (bottom right panel) where $k=10^7$, the oscillations in the NDM spectrum are no longer visible, as their period is too small to be resolved by \deap. The shape of the slope is still distinct from all WIMP spectra, and so can still be distinguished. However this is under the assumption that the WIMP's velocity is described by a Maxwell-Boltzmann distribution. By considering an alternative velocity distribution, a WIMP spectrum could mimic that of NDM's seen here. Past the point where the oscillations are no longer resolved, the shape of NDM spectra is fixed, and so \nreq plateaus, as seen in \Fref{fig:hclfit}.

What one notices across the selection of spectra presented in \Fref{fig:examplespectra} is that the mass of the most indistinguishable WIMP, $m_\text{WIMP}$, approximately follows \nreq. When \nreq is smaller, so is $m_\text{WIMP}$ as the NDM spectra start with sharp declines, which WIMP spectra best mimic by decreasing more quickly as well; this trend is created by decreasing the WIMP mass, which causes the distribution of recoil energies to drop off faster.

Comparing \deap and \xenon in \Fref{fig:hclfit} one notices the range of $k$ values where distinguishing could be possible (before the plateau region and after the divergence at low $k$) is greater for \deap, where \nreq is also smaller. Both these qualities reflect the benefit of having a better energy resolution when searching for NDM. Furthermore the range of distinguishable $k$ values seen in these plots is in good agreement with the those predicted earlier: \Fref{eq:krange} and \Fref{eq:krangeX1T} for \deap and \xenon respectively.

The oscillations in \nreq have a larger period for \xenon, which is counter-intuitive given that the period of the recoil spectra's oscillations is smaller. This is a result of the energy threshold being lower; as $k$ increases the dips in the recoil spectrum contract towards the origin, and so get ``bunched" around it. This means that by having a lower threshold, troughs will pass through this boundary at a slower pace with increasing $k$, and so the period of the oscillations in \nreq increases.

As discussed, the minimum number of events required to distinguish to $3\,\si$ in \deap is 23, which occurs when $k\sim 5000$. For \xenon, the minimum is 55 events (as the energy resolution is worse), which occurs when $k\sim 8000$. As the position of the first trough in the recoil spectrum is at lower $E_R$ in \xenon, one might expect to find this global minimum at a lower $k$ value than in \deap. However because the energy threshold is also lower, the relative differences are such that the minimum is pushed higher.

The final significant difference one notices between detectors, is the size of \nreq in the plateau region, which is lower for \xenon. While the characteristic troughs are not resolved here, the NDM spectra have different slopes. In the case of \xenon, the slope is sharper and more easily distinguished as the target nuclei are larger.

\subsection{Discovery Potential}\label{sec:Discovery}
%
\begin{figure}[ht!]
\centering
\begin{subfigure}[b]{\textwidth}
\centering
\includegraphics[width=0.75\linewidth]{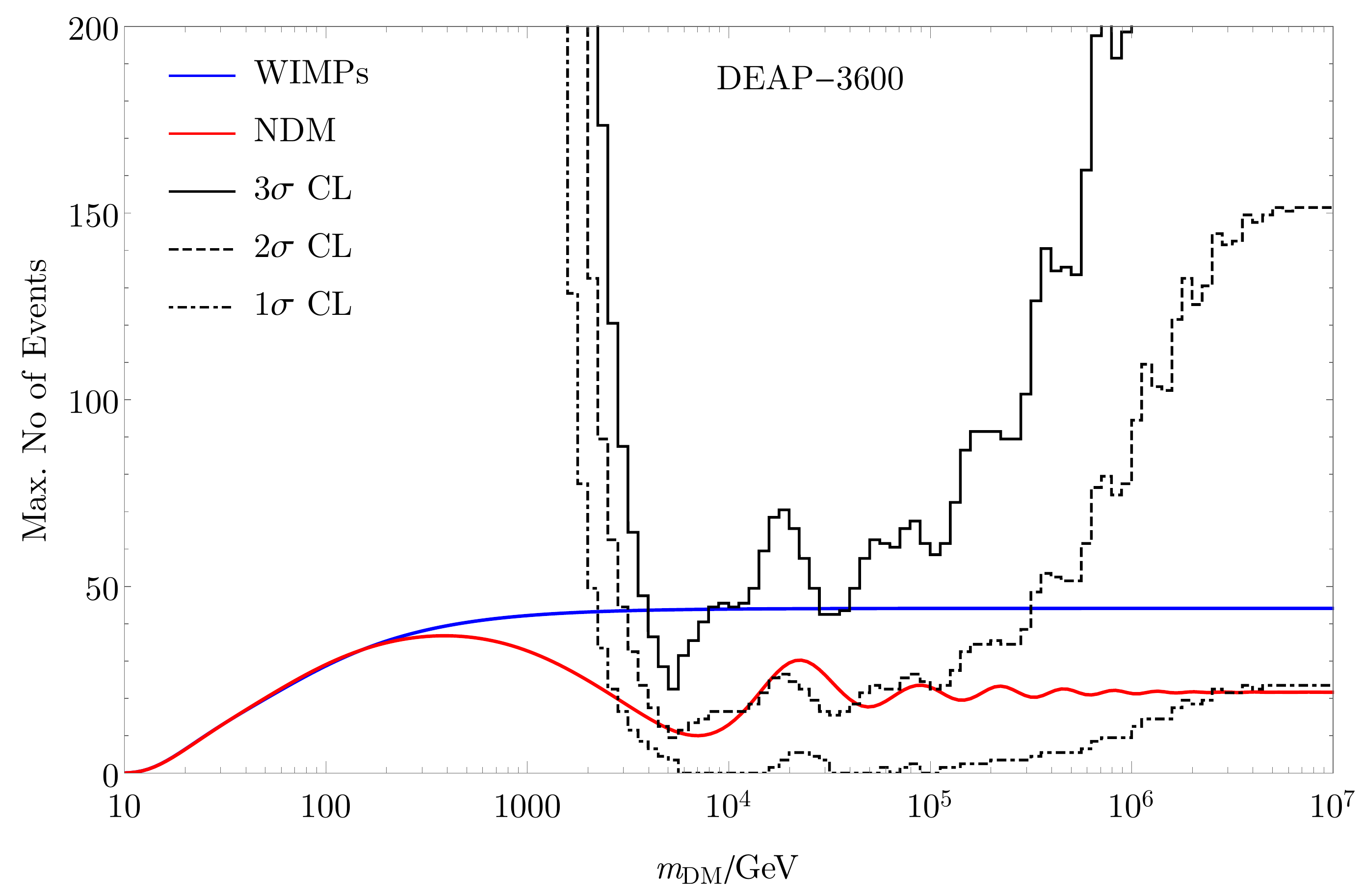}
\end{subfigure}

\begin{subfigure}[b]{\textwidth}
\centering
\includegraphics[width=0.75\linewidth]{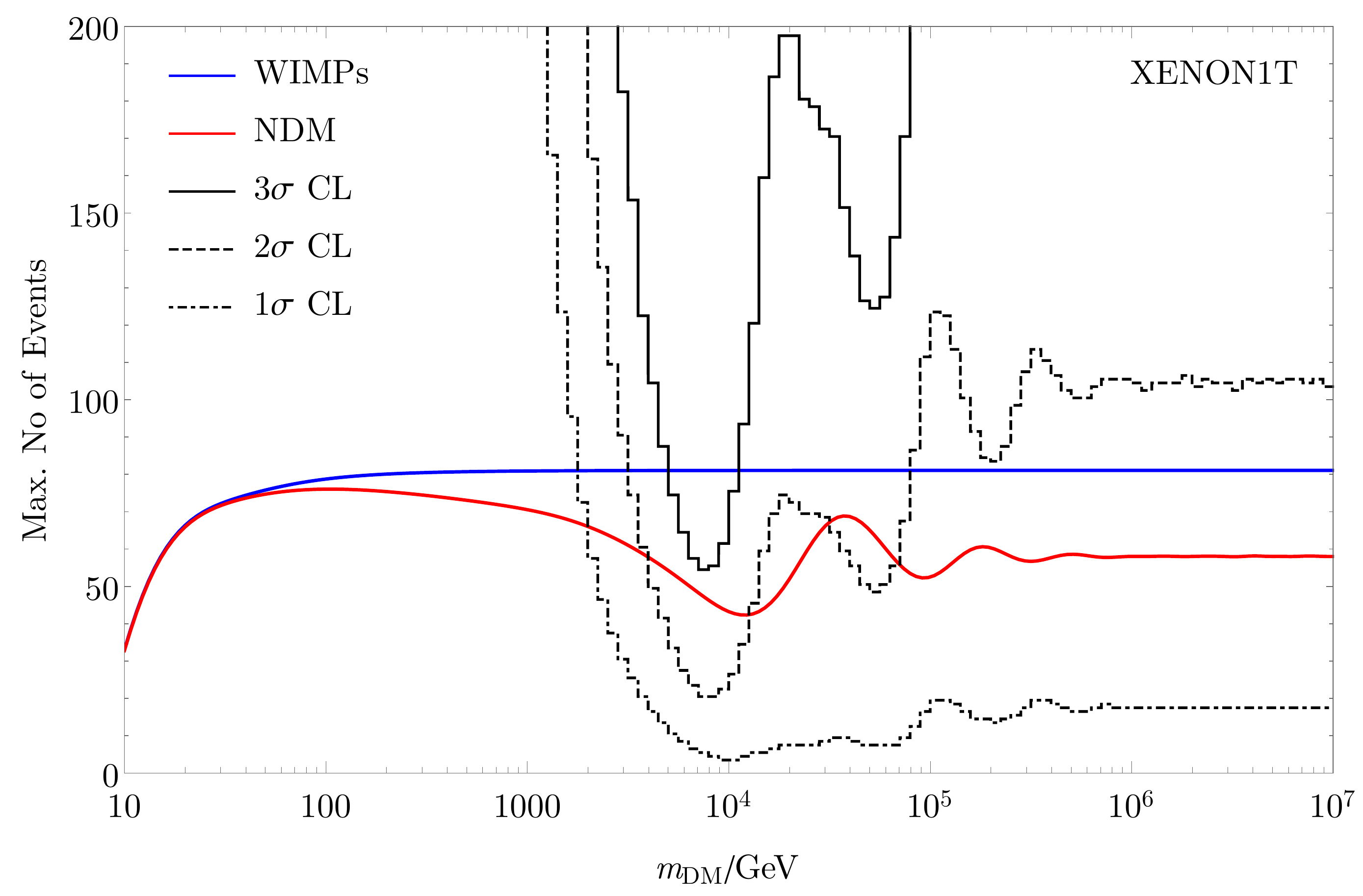}
\end{subfigure}
\caption{The maximum number of events that could be seen at \deap (above) and \xenon (below) against \dm mass for NDM (red) and WIMPs (blue), plotted alongside the 1, 2, \& $3\,\si$ CL lines (the dot-dashed, dashed, and bold lines respectively) for distinction as seen in \Fref{fig:hclfit}.}
\label{fig:Events}
\end{figure}
Finally, we compare the largest number of events (\nmax) that could be seen between NDM and WIMP \dm, in the tonne-scale experiments coming online now. This is estimated by taking the cross-section to be the LUX limit. \Fref{fig:Events} shows \nmax vs. mass for both NDM and WIMPs in the \deap and \xenon experiments, as well as the \nreq values at 1, 2, \& $3\,\si$ CLs. For both WIMPs and NDM, \nmax goes to zero at low mass, where the LUX energy threshold is lower than the relatively conservative projections from \deap and \xenon. The predicted \nmax rises as the experiments increase in sensitivity relative to the current bound for both NDM and WIMPs, however the two models diverge around 100 GeV, where the value plateaus at $\sim 45$ (80) events for WIMPs, but oscillates around $\sim 20$ (60) for NDM in \deap (\xenon). These oscillations occur for the same reason as those in the $\si_0$ limit in \Fref{fig:Limits}; \ie the energy window is enclosing varying configurations of the recoil spectra, which results in a fluctuating \nmax.

Comparing \nmax to \nreq for NDM, we see that in this scenario there are ranges of $m_\text{DM}$ where distinguishing to $1\,\si$ - $2\,\si$ CL could be possible in the current tonne-scale experiments individually. This conclusion is most sensitive to the energy threshold and resolution assumed for each experiment in this study. 

Throughout our analysis we have set $m_1 = 1$ GeV and $R_1 = 1$ fm. If we were to change these values, we do not expect the discovery potential to change, \ie distinguishing to $3\,\si$ or over will still not be possible, at least in the high mass region ($m_k \gtrsim 30$ GeV). Instead we expect changes in $m_1$ and $R_1$ to expand or contract \nreq and \nmax equivalently in \Fref{fig:Events}, and so their values relative to each other do not alter. 

In the high mass region $m_1$ does not effect $g(v_\text{min})$ significantly and so only scales the scattering rate linearly due to the changes in the number density. Since \nreq only depends on the relative shape of WIMP and NDM spectra, it is unchanged by varying $m_1$. Similarly \nmax is not affected as even though the LUX limit changes (by a factor of $m_1$), the scattering rate in \deap and \xenon changes by $1/m_1$. Therefore changes to $m_1$ will not affect the discovery potential in the high mass region. The same can be said of any factor which simply scales the scattering rates.

Varying $R_1$ changes the size of the \kdn state, $R_k$, which affects the scattering rate via the dark form factor, so the configuration of oscillations would alter. However this would be equivalent to varying $k$ as this also changes $R_k$. Varying $k$ also affects $m_k$ (which $R_1$ does not), but as mentioned above, such a  normalisation factor is irrelevant to the discovery potential. Therefore, shifting $R_1$ or $m_1$ will simply stretch or contract \nmax and \nreq equivalently along the mass axis in \Fref{fig:Events}. The situation is more complex in the lower mass region where changes to $g(v_\text{min})$ must also be taken into account.

\begin{figure}[ht!]
\begin{center}
\includegraphics[width=0.75\linewidth]{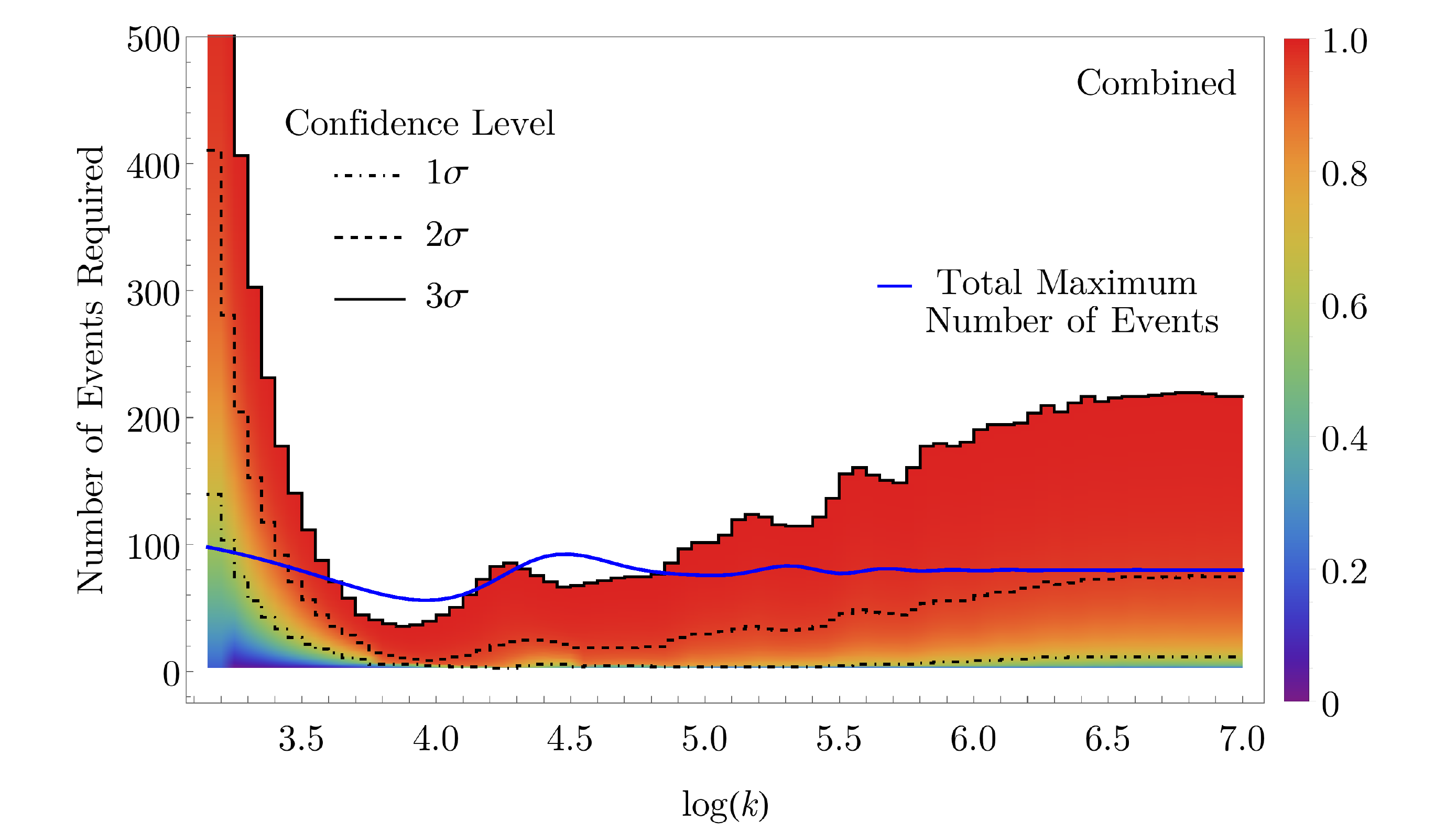}
\caption{The number of events required to positively identify NDM at a given CL with \deap and \xenon combined. The total maximum number of events from both experiments (blue line) reveals two regions of $k$ which can be positively identified as NDM at a $3\,\sigma$ confidence level. }
\label{fig:combined}
\end{center}
\end{figure}

Combining the results from \deap and \xenon would give increased sensitivity to NDM for a range of parameter space, roughly from 3.8 $< \log(k) <$ 4.8. To estimate this, we combine the misidentification probabilities from \Fref{eq:misprob} to find an overall probability $C_{\text{comb}}$ of positively identifying NDM in at least one detector, \ie
\begin{equation}
C_{\text{comb}} = 1-p_{\text{D}}p_{\text{X}}.
\end{equation}
\Fref{fig:combined} shows the result of this for total numbers of events seen across both detectors. The fractional contribution to the total from each experiment is calculated from the expected number of events over each of the detectors' nominal run times. The number of events required is plotted with the total maximum number of events which shows two regions in the range  from 3.8 $< \log(k) <$ 4.8 where there is a $3\,\sigma$ CL of distinguishing NDM from WIMP dark matter.

While $2\,\si$ ``hints" of NDM could be seen in the individual tonne-scale experiments coming online now, and $3\,\si$ level is achievable over a moderate range of parameter space by combining results, larger follow-on detectors under consideration would have significantly enhanced sensitivity. Here we make a simple estimate of the sensitivity for argon (20T) and xenon (7T) targets, since such detectors have been recently proposed or are in the development stage. Taking a 20T argon detector's specifications (the efficiency, etc.) to be the same as \deap's, then \nmax in \Fref{fig:Events} can simply be scaled for the follow-on detector. Thus we can expect \nmax in this detector to oscillate around 400 events with a peak of 800, which is enough to distinguish the entire $k$ range to at least $3\,\si$ where it is possible to do so. Making the same estimate for a 7T xenon detector with the same specifications as \xenon, we can expect \nmax also to oscillate around 400 events with a peak of 650, which similarly distinguishes NDM from WIMPs to at least $3\,\si$. These extrapolations are conservative estimates, as better efficiency, energy thresholds, and so forth, are to be expected.

The outlook for searches for NDM in \deap and \xenon is much the same: distinguishing NDM spectra to $2\,\si$ CL is possible in both, individually. While the number of events required to distinguish is lower in \deap, as it has better energy resolution and oscillations in the recoil spectra have smaller periodicity, \xenon benefits from a reduced lower energy threshold which pushes the sensitivity higher. An ideal detector for probing NDM would optimise for the lowest possible energy resolution, and preferably have a liquid argon target to access the broadest range of $k$ values, as long as the lower energy threshold could be sufficiently small.

\section{Discussion and Conclusion}\label{sec:Con}
Typically \dm models assume \dm is a point-like particle, however composite \dm forming spatially-extended bound states is an area of increasing interest. If \dm is composite, it will have unique phenomenology in direct detection searches, which often provide our most stringent \dm bounds. This paper studies the signals of NDM, following \cite{Hardy:2014mqa}, in the tonne-scale argon and xenon experiments coming online now, \deap and \xenon respectively.

The model considered in this paper is motivated by SM nuclear physics with composite \dm states consisting of DNs bound together under an analogous strong nuclear force. A striking feature of this class of model is the oscillations in the dark form factors, which may be seen in the tonne-scale direct detection searches, when realistic detector energy response, thresholds, and resolutions are taken into account.

We find predicted limits on benchmarks cases of NDM assuming no events are seen by \deap or \xenon, along with those inferred from the null result from LUX. The limits on $\si_0$, the DN-SM nucleon scattering cross-section, are orders of magnitude below the equivalent WIMP limits, owing to a $k^2$ enhancement arising in the scattering rate. The projected limits on WIMP and NDM scattering cross-sections are found to be slightly stronger in \xenon than in \deap. The benefit of argon over xenon targets in the high mass region is reduced as the lower energy threshold is higher and so \deap is probing deeper into the dark form factor where the rate is suppressed. However if \deap's lower energy threshold were lowered to $\sim 8$ keV, then it would be more sensitive than \xenon to NDM in the high mass region.

Likelihood tests were carried out to determine how many events \deap and \xenon would need to detect in order to distinguish between WIMP and NDM spectra. The number of required events for each value of $k$ was found to be dependent on the detector energy threshold, which determines the relative contribution of the second NDM peak in the energy window, and the energy resolution, which determines the amplitude of the characteristic oscillatory features.

The lowest number of events required for the same level of distinction was found to be 23 events in \deap and 55 in \xenon, which occurs for $k \sim 5000$ and $\sim 8000$ respectively. After this point, the required number of events oscillates with $k$ as the configurations of the recoil spectra in the energy window vary. These fluctuations reflect the relative difficulty of resolving the second peak in the energy window, which is the dominant distinguishing feature. 

Setting the DM elastic scattering cross-section to the limit derived from LUX, the maximum number of events which could be seen in \deap and \xenon was derived and found to oscillate around 20 and 60 respectively. While this is likely to be only enough to see ``hints" of NDM in individual detectors, combining results can reach  $3\,\si$ over a range of $k$ values, and future upgrades to both detectors would be able to distinguish NDM from a WIMP to $3\,\si$ over the entire $k$ range considered.

In this analysis we have compared the detection of NDM of a single size to a WIMP with a decaying exponential recoil spectrum. As outlined in the introduction, the choice of NDM model was driven by the unique spectral features produced by the single dark form factor. An extension to this analysis would be to consider the recoil spectrum produced by a distribution over dark nuclei sizes. The resulting recoil shape no longer contains non-decreasing sections and by exploiting the uncertainty in the DM velocity distribution, a WIMP recoil spectrum could easily mimic this NDM spectrum in a single detector. An obvious extension to this work therefore is to consider a multi-target halo-independent analysis to distinguish between NDM with a distribution of nuclear sizes and WIMP DM. 

On the WIMP side we have considered a DM state which generates a decaying exponential recoil spectrum. We note that models of WIMP DM can include interactions that are momentum dependent and this momentum dependence can lead to features in the recoil spectrum, in particular non-decreasing sections at low recoil energies \cite{Chang:2009yt}. Such a spectrum would be much more challenging to distinguish from the NDM spectra considered in this paper and is an analysis that is worth pursuing in future work.


\section*{Acknowledgements}

We acknowledge support from the Science and Technology Facilities Council (grant number ST/L000512/1). 


\bibliographystyle{JHEP}
\bibliography{Bibliography}

\end{document}